\def\wig#1{\mathrel{\hbox{\hbox to 0pt{%
          \lower.5ex\hbox{$\sim$}\hss}\raise.4ex\hbox{$#1$}}}}

\def\etal{{\it et~al.\,}}
\def\mo{$M_\odot$}
\def\lo{$L_\odot\,$}
\def\mj{$M_{\rm J}\,$}

\def\Dwa{$\,$\uppercase\expandafter{\romannumeral5}$\,$}
\def\mic{$\mu$m}

\def\sles{\lower2pt\hbox{$\buildrel {\scriptstyle <}
   \over {\scriptstyle\sim}$}}
\def\sgreat{\lower2pt\hbox{$\buildrel {\scriptstyle >}
   \over {\scriptstyle\sim}$}}
\documentstyle[12pt,aasms4]{article}

\lefthead{Burrows \etal}
\righthead{Non--Gray Models}
\slugcomment{Submitted to the Ap. J.}
\slugcomment{Univeristy of Arizona Theoretical Astrophysics Program preprint 95-26}
\slugcomment{\today}

\begin{document}

\title{A Non-Gray Theory of Extrasolar Giant Planets and Brown Dwarfs}

\author{A. Burrows\altaffilmark{1}, M. Marley\altaffilmark{2}, W.B. Hubbard\altaffilmark{3},  
        J.I. Lunine\altaffilmark{3}, T. Guillot\altaffilmark{4}, D. Saumon\altaffilmark{5}, 
        R. Freedman\altaffilmark{6}, D. Sudarsky\altaffilmark{1}, and C. Sharp\altaffilmark{1}}

\altaffiltext{1}{Department of Astronomy and Steward Observatory, 
                 University of Arizona, Tucson, AZ \ 85721}
\altaffiltext{2}{Department of Astronomy, New Mexico State University, 
                 Box 30001/Dept. 4500, Las Cruces NM 88003}
\altaffiltext{3}{Lunar and Planetary Laboratory, University of Arizona,
                 Tucson, AZ \ 85721}
\altaffiltext{4}{Department of Meteorology, University of Reading, P.O. Box 239,
                 Whiteknights, Reading RG6 6AU, United Kingdom}
\altaffiltext{5}{Department of Physics and Astronomy, Vanderbilt University, Nashville, TN 37235}
\altaffiltext{6}{Sterling Software, NASA Ames Research Center, Moffett Field CA 94035}

\begin{abstract}

We present the results of a new series of non--gray calculations of the atmospheres, spectra, colors,
and evolution of extrasolar giant planets (EGPs) and brown dwarfs for effective temperatures below
1300 K.  This theory encompasses most of the mass/age parameter space occupied by substellar objects
and is the first spectral study down to 100 K.   
These calculations are in aid of the multitude of searches
being conducted or planned around the world for giant planets and brown dwarfs and reveal 
the exotic nature of the class.   Generically, absorption by H$_2$ at longer 
wavelengths and H$_2$O opacity windows at shorter wavelengths conspire to
redistribute flux blueward.  Below 1200 K, methane is the dominant carbon bearing molecule and
is a universal diagnostic feature of EGP and brown dwarf spectra.  
We find that the primary bands in which 
to search are $Z$ ($\sim$1.05 \mic), $J$ ($\sim$1.2 \mic), $H$ ($\sim$1.6 \mic), 
$K$ ($\sim$2.2 \mic), $M$ ($\sim$5 \mic), and $N$ ($\sim$10 \mic), 
that enhancements of the emergent flux over blackbody values, in particular
in the near infrared, can be by
{\it many} orders of magnitude,  and that the infrared colors of EGPs and brown dwarfs are much bluer than previously
believed.  In particular, relative to $J$ and $H$, the $K$ band flux is reduced by CH$_4$ and H$_2$ absorption.  Furthermore, we derive  
that for T$_{\rm eff}$s below 1200 K most or all true metals are sequestered below the photosphere,    
that an interior radiative zone is a generic feature of
substellar objects, and that clouds of H$_2$O and NH$_3$ are formed for T$_{\rm eff}$s
below $\sim$400 K and $\sim$200 K, respectively.
This study is done for solar--metallicity objects in isolation and does not include the effects of stellar insolation.
Nevertheless, it is a comprehensive attempt to bridge the gap between the planetary and stellar realms 
and to develop a non--gray theory of objects from 0.3 \mj (``saturn'') to 70 \mj ($\sim$0.07 \mo).
We find that the detection ranges for brown dwarf/EGP discovery of both ground-- and space--based telescopes  
are larger than previously estimated.

\end{abstract}



\section{Introduction}

After years of slow progress and ambiguous, but tantalizing, observations of objects in the
field and in young clusters, the study of brown dwarfs via reflex stellar motion, filter photometry,
and spectroscopy has finally come into its own.
The direct detection of Gl229B
(Oppenheimer \etal\  1995; Nakajima \etal\ 1995; Matthews \etal\ 1996; Geballe \etal\ 1996; 
Marley \etal\ 1996; Allard \etal\ 1996; Tsuji \etal\ 1996) was a watershed because Gl229B displays methane spectral features
and low surface fluxes that are unique to objects with effective temperatures (in this case, T$_{\rm eff}$$\sim$950 K)
below those at the solar--metallicity main sequence edge ($\sim$1750 K, Burrows \etal\ 1993).
In addition, the almost complete absence of spectral signatures of metal oxides and hydrides (such as
TiO, VO, FeH, and CaH) is in keeping with theoretical predictions that these species are depleted 
in the atmospheres of all but the youngest (hence, hottest) substellar objects 
and are sequestered in condensed form below the photosphere (Lunine \etal\ 1989; Marley \etal\ 1996).
This remarkable convergence between theory and observation should not obscure the fact that
the study of the atmospheres, spectra, colors, and evolution of substellar objects is still in its
infancy.  Though predictions of luminosity, T$_{\rm eff}$, and radius evolution versus mass and composition
have been available for almost a decade (Nelson \etal\ 1985; 
D'Antona \& Mazzitelli 1985; Burrows, Hubbard, \& Lunine 1989; 
Dorman \etal\ 1989; Stevenson 1991; Stringfellow 1991; Burrows \etal\ 1993) 
and predictions for the colors and spectra of hot ($\sgreat$ 2000 K), young brown dwarfs
have been available for a few years (Allard \& Hauschildt 1995), to date there has been no {\it complete} theory of the evolution
of the colors, spectra, and structure of brown dwarfs with temperatures below $\sim$1300 K.  This is true despite 
the fact that, for most of the mass--age space occupied by brown dwarfs in the galaxy, T$_{\rm eff}$ is indeed below 1300 K.
To remedy this situation, we here present the first non--gray theory of the evolution, spectra,
and colors of brown dwarfs down to T$_{\rm eff}$s of 100 K.  

This sweep of T$_{\rm eff}$s and the effective physical equivalence between  
equal--mass extrasolar giant planets (EGPs) and brown dwarfs allows our theory to apply, without modification,
to EGPs as well.  It is sensible to distinguish EGPs and brown dwarfs on the basis of their origins:
brown dwarfs form like stars, but are too light to burn hydrogen stably on the main sequence, 
and EGPs form out of protoplanetary disks by accretion.  The different birth paths no doubt lead to
different metallicities, rotation rates, and orbital characteristics, and 
in the giant planet case to the presence of an ``ice/rock'' core (Podolak, Hubbard, \& Pollard 1993).
However, in the main it is not the origin, but the mass, composition, age, and proximity of a 
hydrogen--dominated object to a star that determines its spectral signatures and
evolution.  An object's pedigree is not an obvious observable.   

\renewcommand{\thefootnote}{\fnsymbol{footnote}}
\setcounter{footnote}{1}

We have already published a general theory of extrasolar giant planets with masses from 0.3 \mj
to 15 \mj, where \mj denotes a Jupiter mass ($\sim$0.001 \mo)  (Burrows \etal\ 1995; Saumon \etal\ 1996; Guillot \etal\ 1996).  
When we published these EGP models, no such
objects had been identified.  Now, Doppler spectroscopy alone has revealed about 20 objects
in the giant planet/brown dwarf regime, including companions to $\tau$ Boo, 51 Peg, $\upsilon$ And,
55 Cnc, $\rho$ CrB, 70 Vir, 16 Cyg, and 47 UMa (Butler \etal\ 1997; Cochran \etal\ 1997; 
Marcy \& Butler 1996; Butler \& Marcy 1996; Mayor \& Queloz 1995;
Latham \etal\ 1989).
However, our old theory 
assumed that EGPs emit like blackbodies.  While this assumption is not bad for some wavelength stretches in the 
mid--infrared, it can be spectacularly off in bands of interest for direct detection.  The Gl229B
campaign taught us that.  The non--gray theory we develop 
in this paper encompasses masses from 0.3 \mj to 70 \mj$\!$\footnote{
For a 70 \mj\ brown dwarf, T$_{\rm eff}$ $<$ 1300 K for $\sgreat\ $4 Gyr.} and is in aid of the multitude of direct searches
for substellar objects, be they ``planets'' or brown dwarfs, upon which the world's astronomical community 
has now collectively embarked (TOPS and ExNPS reports; Leger \etal\ 1993).  In order to maintain a reasonable focus, we 
limit ourselves to solar--metallicity (Anders \& Grevesse 1989) objects in isolation and ignore the
effects of stellar insolation (Guillot \etal\ 1996).  A subsequent paper will address the consequences of proximity
to a central star and of varying metallicity.

In \S 2, we describe our calculational techniques and the opacity and thermodynamic data that we
used.  Section 3 covers the physics of the atmospheres of brown dwarfs and EGPs and includes a discussion
of temperature/pressure/composition profiles and the location of convective and radiative zones.
In \S 4, we describe the evolution of objects with masses from 0.3 \mj to 200 \mj, from saturns
to M dwarfs, and provide a global view of the giant planet/brown dwarf/M dwarf model continuum. 
Section 5 contains a comprehensive discussion of the near-- and mid--infrared spectra of EGPs and brown
dwarfs as a function of mass and age, as well as T$_{\rm eff}$ and gravity.  Our major results 
are to be found in this section.  Section 6 presents the IR colors and magnitudes from the $J$ through the $N$ 
bands and demonstrates just how unlike blackbodies these objects can be.  Search and discovery 
techniques using Doppler spectroscopy,
astrometry, transits, and microlensing are rapidly maturing, but it is only via direct photometric
and spectroscopic characterization that substellar objects will really be understood.  In \S 7, we 
summarize the salient features of the non--gray theory and list some of the  
ground-- and space--based telescopes and searches for which 
it should prove useful.

\section{Input Physics}

The ingredients for a theory of EGPs and brown dwarfs include (1) equations of state for 
metallic hydrogen/helium mixtures and molecular atmospheres, (2) chemical equilibrium
codes and thermodynamic data to determine the molecular fractions, (3) scattering and absorption opacities for
the dominant chemical species, (4) an atmosphere code to calculate temperature/pressure
profiles and to identify the radiative and convective zones, (5) an algorithm for converting a grid
of atmospheres into boundary conditions for evolutionary calculations,
(6) a Henyey code, and (7) a radiative transfer code to provide emergent
spectra.  In principle, the calculation of the atmosphere, involving as it does radiative transfer,
and the calculation of the emergent spectrum are done together.  However, as long as the
thermal profiles obtained with the atmosphere code are accurate, one can 
employ these profiles, but with another more accurate and higher--resolution transfer scheme, 
to calculate emergent spectra.  Though we use the k--coefficient method to calculate the
atmosphere profiles, we are free to employ other radiative transfer schemes to obtain 
spectra.    

In the evolutionary calculations, we use the Saumon \& Chabrier (1991, 1992) equation of state in the metallic and high--density
molecular regimes.  For solar metallicity, 
near and above brown dwarf/EGP photospheres, throughout most of their lives 
the dominant equilibrium form of carbon is CH$_4$, not CO (Fegley \& Lodders 1996), that 
of oxygen is H$_2$O, and that of nitrogen is either N$_2$ or NH$_3$, depending upon T$_{\rm eff}$.  
Hydrogen is predominantly in the form
of H$_2$.  Silicates and metals are found at high optical depths and temperatures.
Clouds of NH$_3$ and H$_2$O can form for T$_{\rm eff}$s below
$\sim$200 K and  $\sim$400 K, respectively.  While for this new model suite we have precipitated species according
to their condensation curves, we have not consistently incorporated the effects of the associated 
clouds.  If a species has condensed, it is left at its saturated vapor pressure.  
Though the proper inclusion of the radiative transfer effects of clouds is deferred to a later work, in \S 2.6 \& \S 3 we discuss the
basic physics of such clouds for T$_{\rm eff}$s between 100 K and 1300 K (see also 
Lunine \etal\ 1989) and speculate on their role in spectrum formation.

\subsection{Opacities}

Water is an important source of opacity in EGPs and brown dwarfs, particularly when the
many lines that originate from highly excited energy levels are considered.  
A series of databases have recently become available that are based
on theoretical calculations that employ a variety of quantum mechanical methods
(Polyansky, Jensen, \& Tennyson 1994; Wattson \& Rothman 1992; Partridge \& Schwenke 1997).   
In particular, Partridge \& Schwenke have calculated the potential energy surface and dipole moment
function using an ab initio method. This energy
surface was empirically adjusted to improve the fit between predictions
and the HITRAN 92 database (Rothman \etal\ 1992) for J $\sles$ 5.  The overall accuracy in wavenumber and
intensity is good and the data have already been used to identify previously
unidentified sunspot lines as water lines (Carbon \& Goorvitch 1994).
The 10, 25, 50, 75, and 90 percentiles of the errors in the line
positions as compared to HITRAN are: -0.11, -0.04, -0.01, 0.02, and 0.07 cm$^{-1}$.
This database allows the inclusion of many predicted lines that are
unobserved in the lab and only become important at higher temperatures, since 
they arise from highly excited levels.
This can be particularly important in opacity windows, {\it i.e.,} regions where the
water opacity reaches a local minimum, but where many weak, high--excitation
lines may occur.  Depending upon the temperature of the layer, and the assumed
abundance of water, well over $2.0\times10^{8}$ lines could be required for a calculation
at the highest temperatures, while far fewer lines are needed at lower temperatures.
We use this new Partridge \& Schwenke H$_2$O database. (Note that the model published by Marley \etal\ (1996)  
used an earlier version of the Schwenke data with
fewer lines.)

For other than H$_2$O, we used databases that go beyond what is
available in the HITRAN database (Rothman \etal\ 1997).  The limitations of 
HITRAN are a consequence of the cutoff in
line strength that is imposed at a temperature of 296 K.  Weaker lines are excluded
by this cutoff, though they may become
much stronger as the temperature climbs to well above 
300 K. In addition, in HITRAN lines whose analysis
is lower in quality may have been excluded.  Since in our application we are
interested in the total opacity only over fairly broad regions in wavenumber,
we can accept lines whose positions and strengths may not be known with the
highest precision.  This makes available databases with far more lines than HITRAN.
The GEISA database (Husson \etal\ 1997)
can also be a source of additional lines, as its line strength cutoff is lower
than that of HITRAN.  Additional lines have been obtained from theoretical
calculations (Tyuterev \etal\ 1994; Goorvitch 1994; Tipping 1990; Wattson \& Rothman 1992) and 
from other researchers prior to publication (L. R. Brown, private communication).  This has resulted in databases
for CH$_4$ and CH$_3$D of $1.9\times10^6$ lines, for CO of $99,000$ lines, for NH$_3$ of $11,400$
lines, for PH$_3$ of $11,240$ lines, and for H$_2$S of $179,000$ lines.

Modeled continuum opacity sources include
$\rm H^-$
and $\rm H_2^-$ opacity
and collision--induced absorption (CIA)  
of H$_2$ and helium (Borysow \& Frommhold 1990; Zheng \& Borysow 1995).  The latter is a direct function of pressure 
and a major process in EGP/brown dwarf atmospheres.
We employed the formulation of Rages \etal\ (1991) for Rayleigh scattering, important at shorter wavelengths.

\subsection{Line Profiles}

For our calculations, when the data are available, the line widths for the 
various molecules are assumed to be due to H$_2$ or
H$_2$ + He broadening. Currently, such data are
available for H$_2$O (Brown \& Plymate 1996; Gamache, Lynch, \& Brown 1996), CO (Bulanin \etal\ 1984; Lemoal \& Severin 1986),
CH$_4$ (Margolis 1993,1996; L. R. Brown, private communication), PH$_3$ (Levy, Lacome, \& Tarrago 1994),
and NH$_3$ (Brown \& Peterson 1994).  These data were derived
from laboratory measurements, and the width for each line was derived from
a fit to the available data as a function of the value of J and the other
relevant rotational quantum numbers. Because these measurements typically 
cover only one or more vibrational bands, it is necessary to assume that any
vibrational dependence of the widths is very small, since the same fit is used for
all bands. 
For other species it is necessary to use the
available N$_2$ widths, under the assumption
that the broadening due to H$_2$ is larger than that due to N$_2$, because of 
the difference in quadrupole moments. This
is still quite uncertain and may vary from species to species.

\subsection{Calculation of the Atmosphere Structures and Spectra}

We construct model atmospheres employing an
approach similar to that used to derive temperature
profiles for the outer planets (Marley \etal\ 1997)
and Titan (McKay \etal\ 1989).  
The model consists of a series of up to 60 homogeneous plane--parallel
layers.  The bottom of the lowermost layer is placed at a depth
of up to 300 bars,
the top of the uppermost at $0.5\,\rm mbars$.  Levels, which
separate layers, are spaced approximately evenly in $\log P$. 
The current model considers 101 spectral intervals from
$0.87\,\rm \mu m$ to $2.5\,\rm cm$.

A trial temperature profile is adjusted until the entire atmosphere
is in radiative equilibrium.   
The model is judged to be in radiative equilibrium when the
temperature profile results in zero net flux across each level in
the radiative region of the atmosphere. Layers in which the
radiative lapse rate exceeds the adiabatic lapse
rate are then deemed convective.  The lapse rate is adjusted
only one layer at a time and a new radiative--convective equilibrium
profile is computed after the lapse rate in each layer
is adjusted.  This approach allows for the presence of multiple
convection zones and works very well for the solar
Jovian planets (Marley \etal\ 1997).  
The scaling
arguments of Ingersoll \& Pollard (1982) suggest that the
fractional difference between the actual temperature profile and
an adiabat is less than $10^{-4}$ for Jovian atmospheres.

Fluxes are computed using the two--stream source function technique (Toon
\etal\ 1989). The intensity is integrated over
five angular Gauss points in each hemisphere.  This technique is rapid and is well suited
for application to inhomogeneous, multiple--scattering atmospheres.
Toon \etal\ discuss the accuracy of this
approach for a variety of scattering and non--scattering cases.
For the case of no scattering, they found layer emissivities
to agree with an exact calculation to better than 1\% over a three
decade range of layer optical depths.

\subsection{The k--coefficient Method}

The k--coefficient
method (Goody \etal\ 1989; Lacis \& Oinas 1991) is widely used in
planetary atmosphere modeling.  This is not the ODF technique (Saxner \& Gustafsson 1984)
and gives excellent
agreement with full line--by--line computations of atmospheric transmission.
Typical errors are between 1 and 10\%
(Grossman \& Grant 1992,1994a,1994b).

After we perform the line--by--line calculations for each individual molecule, the resulting
opacity files are put on a common frequency scale.  Then, they are combined into one
file for the total opacity using the appropriate mixing ratios
for each molecule.  
The abundance of the various constituents is determined by a
chemical equilibrium calculation and condensation (\S 2.5 \& \S 2.6).  At temperatures
below the condensation point of a given constituent, the abundance
is set equal to the saturation vapor pressure. 
In the k--coefficient method, this summed file is then
broken up into a large number ($\sim$100) of frequency blocks covering the entire
spectrum.  
The creation of the block structure allows one to avoid
solving the transfer equation at a very large number of individual frequency
points (Goody \etal\ 1989).  

One constructs a
distribution function of all the opacities within each frequency block.
This technique
takes all the opacities (at each wavenumber) from the line--by--line calculation,
sorts them according to value (independent of wavenumber), and places them
in logarithmically--spaced bins.  Note that this destroys the relationship
between a particular value of the opacity and the wavenumber where it
occurred.  
This distribution (number of points that fall in each bin versus opacity) is then
normalized and converted into a cumulative distribution.
The distribution is now the cumulative number of points at
each bin versus the opacity of that bin.  After the proper normalization, this
now gives the cumulative probability versus opacity, {\it i.e.} the probability that
an opacity is less than or equal to a given value.  This distribution is then inverted using
interpolation to give the final cumulative probability distribution.
The abscissa is now the probability normalized
to run from 0 to 1.0 and the ordinate is opacity and runs from the minimum to
the maximum value of the opacity within the window.

Because the final distribution is now uniform in probability space, it is
possible to integrate this distribution to find the total opacity in each
window.  We perform this integration 
by extracting values at appropriate Gauss points.  These values, sometimes called
correlated k's, can be used with the appropriate Gauss weights to represent the
total opacity in a given window.  

All molecular absorption is treated within each spectral interval by a weighted
sum of exponentials.  The transmission, $T$, as a function of
absorber column mass, $u$, within a layer for a given spectral
interval is expressed as
$$T(u)=\sum_i w_i e^{-k_iu}\eqno(1)$$
where the $w_i$ are weights and the $k_i$ are equivalent
monochromatic absorption coefficients. 

Within each spectral interval the equation of radiative transfer
is solved eight times as a monochromatic equation of
transfer, once for each term in the expansion in eq. (1), with
the Rayleigh optical properties identical for each
of the computations.  The fluxes thus obtained are then summed,
with the terms weighted by the $w_i$, to obtain the total flux in
the spectral interval.  

The accuracy of
the approach is controlled by the width of the model spectral intervals
and the number of Gauss points used.  To improve the sensitivity to
small fractions of very strong lines in a given spectral interval,
we employ a double Gaussian quadrature.  The first four
Gauss points provide for the integral over the weakest 88\% of the 
lines. The remaining four points take the integration over
the remaining probability interval.
Accuracy can be compromised if the
spectral shape of the molecular opacity within a given interval changes
with depth in the model, usually because of conversion of one species
to another.  Thus, we have carefully chosen the individual spectral intervals
to minimize such effects.  Experimentation with the current spectral intervals
reveals that computed temperature profiles are not sensitive to a further
increase in their number.

During model runs, the k--coefficients at an arbitrary temperature--pressure
point are computed by interpolation within the k--coefficient grid.
Opacities are interpolated in $\log P$ - $1/T$ space to follow the
variation of constituents along the vapor pressure curve.

\subsection{Chemical Equilibrium Calculations}

For this paper, the chemical equilibrium calculations were performed with the ATLAS code and
data from Kurucz (1970). The Kurucz reaction constants are inaccurate at
low temperatures, but, fortunately, the NH$_3$ $\rightarrow$ N$_2$ and CH$_4$ $\rightarrow$ CO
conversions that occur in EGPs and brown dwarfs do so in regions of $T-P$ space for which the
Kurucz reaction constants are accurate.
Condensation of CH$_4$, NH$_3$, H$_2$O, Fe, and
MgSiO$_3$ was included using data from various sources, including Eisenberg \& Kauzmann (1969), the
Handbook of Chemistry and Physics (1993), and Lange's Handbook of Chemistry
(1979). Following Fegley \& Lodders (1994, 1996), we assumed that Al, Ca,
Ti and V were removed either by condensation or were dissolved in silicate
grains at about the MgSiO$_3$ condensation temperature.
These atoms are important because they lead to molecules that are strong
light absorbers, such as TiO and VO. However, they have not been
detected in the giant planets of our solar system and shouldn't be
present in relatively cool objects such as the brown dwarf Gl229B
(Marley \etal\ 1996).
Our results are in excellent agreement with those of Fegley
\& Lodders (1994, 1996).

\subsection{Condensation Processes}

The principal effect of cloud formation is the removal of molecular
species from the gas phase into the solid or liquid phase. For a
single component system that does not interact chemically with
other species ({\it i.e.}, water) cloud formation occurs at a pressure
level (``cloud base'') where the partial pressure of the gaseous
species just exceeds its saturation vapor pressure. The saturation
vapor pressure is given approximately by the Clausius--Clapeyron
relationship, $P=Be^{-A/T}$, where B and A are weak functions of
temperature. A is the latent heat of condensation (per mole) divided by the
universal gas constant. At lower pressures (higher altitudes) in
the atmosphere, the abundance of the gas phase species drops off in
this inverse--exponential fashion. Since A/T $>$ 1, the falloff is
steep and the gaseous species
does not contribute to the emergent spectral distribution at
pressures much above the cloud base.

There are several complications to the above simple picture. The
first is supersaturation. Because the radius of curvature of cloud
particles or droplets adds an additional surface energy to the
condensed phase, the cloud base is usually elevated above the
pressure level at which the ratio of partial pressure to saturation
vapor pressure is unity (this is the supersaturation ratio). In the
case of terrestrial clouds, the threshold ratio may be 1.2; in
certain cold environments lacking particulates to serve as
nucleation sites, the value may be as high as 2 (such as in Neptune; Moses, Allen, \& Yung
1992). Because of the high temperatures at the unity optical depth
level in objects such as Gl229B, we ignore supersaturation; variations of a
few tens of percent in the supersaturation are not discernable in
its spectra.

The second complication is that for most condensable species the
temperature variation of A cannot be ignored in two cases: near the
liquid--vapor critical point, and over large ranges of temperature.
Because our models span a large temperature range, this is a
concern for the major cloud--forming species, in particular water.
For water, we use
multi--term polynomial and exponential fits for the liquid and solid phases
from Eisenberg \& Kauzmann (1969).  These are suitable for temperatures
from well below the ice point to the critical temperature.

The third complication is that many cloud--forming elements are not
incorporated in the same molecular species in the gaseous and
condensed phases, but instead a chemical reaction occurs associated
with the condensation process. To correctly characterize
cloud formation requires that we incorporate it into the
chemical equilibrium computations described above; cloud formation
occurs for particular elements when the phase with the minimum
Gibbs free energy is the condensed phase. The only species we
consider here for which this is an issue is the magnesium silicate, MgSiO$_3$.
Convenient expressions for the resulting condensation are given in Barshay
\& Lewis (1976). We are currently preparing a more
comprehensive set of condensation curves for minor species using a
Gibbs energy minimization routine, the results of which will be presented in a subsequent paper.

\subsection{T$_{\rm eff}$ versus T$_{10}$}

As in our previous papers, we parametrize the specific entropy
of the fully convective deep interior of an EGP by T$_{10}$,
which is the temperature that the interior isentrope would have
if extrapolated to a pressure of 10 bars, and which may or may not
equal the actual atmospheric temperature at 10 bars.

We construct a $T_{\rm eff}-T_{10}-g$ surface by making a splined
fit to  the grid of non--gray model atmospheres
augmented with additional $T_{\rm eff}-T_{10}-g$ points from gray model
atmospheres, as presented in Saumon \etal\ (1996).  The resulting surface
is shown in Figure 1.  The points used to define this surface are shown as
dots.  Open dots represent the previous (gray--atmosphere) relations from
Saumon \etal , and the solid dots are the new (non--gray--atmosphere)
calculations.  To illustrate the domain actually
traversed by EGP models, three evolutionary trajectories of EGPs
with masses of 1 \mj (left), 10 \mj (middle), and 42 \mj
(0.04 \mo; right) are shown.  The earliest portions of the
evolution tracks are shown as dotted lines starting at
an age of $10^{-3}$ Gyr and ending at an age of
$10^{-1}$ Gyr.  Subsequent evolution is shown as a solid line,
which ends at a maximum age of 10 Gyr for the 1 \mj and 10 \mj
models, and at 20 Gyr for the 42 \mj model.  Evolution of objects with ages
greater than 0.1 Gyr is well constrained by our non--gray grid.

\section{Atmospheres}

\subsection{Radiative and Convective zones}

Figure 2 shows profiles of atmospheric pressure as a function of temperature
calculated in our grid of non--gray models.  In this figure, the surface gravity
$g$ is held constant at 2200 cm s$^{-2}$ (close to the value for Jupiter), and
T$_{\rm eff}$ takes the values 128 K (lowest curve), and 200 K to 1000 K
(highest curve) in steps of 100 K.  This sequence does
not precisely represent an evolutionary sequence for a Jupiter--mass object
because $g$ actually decreases with increasing T$_{\rm eff}$ in such an object.
The heavy dot represents the photosphere, which is not a well--defined
region in a non--gray model atmosphere, but which we approximate as the
region in the atmosphere where the local temperature T = T$_{\rm eff}$.
Convection zones, where the local value of $dT/dP$ is essentially
equal to the adiabatic value, are shown as dashed lines; as usual,
radiative zones appear where the local value of $dT/dP$ is
subadiabatic.  Various chemical
boundaries, discussed below, are shown as lighter lines, solid for a
change in equilibrium for chemical species, and dashed for formation
of condensed phases of a single species.  The observed $P-T$ relation
for Jupiter (Lindal 1992; Seiff \etal\ 1996) is shown as a dot--dashed line.

Note in Figure 2 that a detached radiative zone appears in the hotter
models at temperatures around 1500 to 2000 K.  The physical origin
of this zone is the near--coincidence of a minimum in the CIA opacity as
a function of wavelength with the maximum of the local Planck
function, as originally discovered by Guillot \etal\ (1994).  Guillot \etal\ have 
determined that a detached radiative zone is likely also to be present
in Jupiter at temperatures between 1200 and 2900 K.  Not only is
the detached radiative zone of interest in its own right, but it is
important for the evolution of EGPs because it causes the specific
entropy in the uppermost convection zone to be higher than the
specific entropy in the deepest convection zone.  Thus, an EGP in
which this zone appears will evolve slightly more rapidly than it
would in the absence of the zone.  That is, for a given value of
T$_{\rm eff}$, the central temperature will be lower than would be
calculated without allowing for the detached radiative zone.

The models shown in Figure 2 do not extend to sufficient depth 
at T$_{\rm eff}= 128$ K and 200 K to include the detached radiative
zone.  Thus, the relation between T$_{\rm eff}$ and T$_{10}$ for
these models is slightly incorrect.  However, as discussed by Guillot \etal\ (1994) and Ingersoll \& Pollard (1982),
the difference between the radiative gradient and the adiabatic gradient
in the radiative zone is small, so that the cumulative error in
calculating T$_{10}$ is at the level of or smaller than other effects which
we neglect in this paper, such as insolation from a moderately
distant ($\sim 5$ A.U.) companion star.

Figures 3--6 portray the calculated atmospheric profiles for surface
gravities of $10^4$, $3 \times 10^4$, $10^5$, and $3 \times 10^5$
cm s$^{-2}$.  As the gravity increases, the photospheric pressure increases, but
the T$_{\rm eff}$ dependence of the photospheric pressure is weak.
For all gravities, an extended or second radiative zone is a generic feature of the atmospheres.
This will have consequences for the mixing of non--equilibrium species into 
the observable layers.


\subsection{Equilibrium and Condensation Lines}

Figures 2--6 depict temperature/pressure profiles, on
which are superimposed the equilibrium condensation lines for various species.
For Jovian--type effective temperatures condensation of ammonia and
water occur near the photosphere. 
Even for objects as warm
as T$_{\rm eff}$=500 K, water condensation occurs, but it does so at
altitudes well above the photosphere in the atmosphere, and at
pressures so low that (a) the actual cloud particle density is rather
small and (b) the cloud particles are expected to fall out of the
atmosphere rapidly. We therefore expect such a tenuous water cloud
to play a much more minor role in the radiative balance and spectral
appearance than in the cooler objects for which the water cloud is
potentially quite massive (as in simplistic Jupiter models).

Interestingly, among cloud--forming species which are abundant by
virtue of cosmic composition, a relatively large gap occurs between
water and less volatile species. Sulfur--bearing condensates of iron
sulfides (not shown in the figures) might be present in the effective
temperature range around 500 K. Beyond that, magnesium silicate and
iron clouds are expected to form around the photosphere for objects
with effective temperature exceeding 1000 K. All of the relationships
between effective temperature and cloud formation are modestly
sensitive to the effective gravity.

Also shown in Figures 2--6 are lines defining equal gas--phase abundances
of methane and carbon monoxide and of ammonia and nitrogen. 
Le Chatelier's principle demands that
the hydrate species dominate at the lower temperatures. Hence, below
$\sim$1200 K and $\sim$600 K, methane and ammonia, respectively, are the dominant carbon
and nitrogen species.  Between $\sim$600 K and $\sim$1200 K,
N$_2$ and CH$_4$ can coexist.
This illustrates that Gl229B is a threshold object which
may contain some amounts of CO in addition to CH$_4$. It is also
possible that the atmosphere of Gl229B contains detectable amounts of ammonia, because,
even though it is a minor species, it is spectroscopically active.

\subsection{Clouds}

Cloud formation depletes a gas--phase
absorber from certain regions of the atmosphere; if this occurs
around the photosphere the resulting radiative balance and emergent
flux distribution are modified.  Because of condensation, we expect that 
the gaseous water bands will disappear for objects with effective
temperature below about 400 K.  We
expect the disappearance of silicate or iron features below about
1000 K (depending modestly on surface gravity).

Beyond predicting where the water and ammonia bands should
disappear due to condensation, the spectral and radiative effects
of clouds are difficult to quantify. Simple models in which clouds
are uniformly distributed over the surface of the EGP, and are
characterized by a single particle size, fail to take account of
atmospheric dynamics which can lead to dramatic changes in the
effects of clouds. In particular, convective processes lead to
growth in the mean particle size, as well as a potentially
heterogenous distribution of clouds across the disk of the object.
In the case of water and magnesium silicates, the latent heat of
condensation increases the mean upwelling velocity and can
exaggerate these effects, as quantified by Lunine \etal\ (1989).
The simple model of the transport processes in magnesium silicate
clouds presented in Lunine \etal\ suggests particle sizes in the range of
100 microns are possible by coalescence, much larger than the
micron--sized particles one would assume from simple condensation. The
radiative properties of a cloud clearly depend upon the
actual particle size, as well as the large--scale cloud
morphology (broken or continuous).

The importance of these processes is seen in Jupiter. Earth--based
and Voyager spectra, along with theoretical modeling, show
that the spectroscopic effects of water
clouds differ from those predicted by the simplest condensation
models (Carlson \etal\ 1987). Galileo probe results (Niemann \etal\ 1996)
demonstrate directly that global dynamical processes combined with
condensation lead to a strongly heterogeneous distribution of water
clouds across Jupiter's disk.  Thus, in the archtypical example of a
giant planet, the simple assumptions about cloud
formation and their impact on radiative processes fail.
Likewise, on Neptune
the methane clouds are distributed in a manifestly heterogeneous fashion.

For these reasons we have chosen not to model explicitely the
spectral and radiative effects of condensed species. To do so with
the available information remains an unconstrained exercise, but
higher resolution spectra on objects such as Gl229B could
provide constraints for such cloud modeling.

\section{Evolutionary Models}

In Burrows \etal (1995) and Saumon \etal (1996), we published cooling curves for
EGPs and small brown dwarfs that were based upon our then--current atmosphere models.
For this paper, we have updated the H$_2$ CIA, H$_2$O, and CH$_4$ opacities and the T$_{10}$--T$_{\rm eff}$
grid.  Consequently, the evolutionary tracks have changed, but generally by no more than 10\%
in luminosity at any given time, for any given mass.  In this section,  we present these latest cooling tracks
and do so in the larger context of the M dwarf/brown dwarf/EGP continuum.   The figures in this section
cover three orders of magnitude in mass and encapsulate the characteristics of the 
entire family of substellar objects and the transition to stars. 

Figure 7 portrays the luminosity versus time for objects from Saturn's mass (0.3 \mj) to 0.2 \mo.
The early plateaux between 10$^6$ years and 10$^8$ years are due to deuterium burning, where 
the initial deuterium mass fraction was taken to be 2$\times$10$^{-5}$.  Deuterium burning occurs earlier,
is quicker, and is at higher luminosity for the more massive models, but can take as long
as 10$^{8}$ years for a 15 \mj object.  The mass below which less than 50\% of the ``primordial''
deuterium is burnt is $\sim$13 \mj (Burrows \etal\ 1995).  On this and subsequent figures in this section, we have arbitrarily
classed as ``planets'' those objects that do not burn deuterium and as ``brown dwarfs'' those that do burn deuterium,
but not light hydrogen.  While this distinction is physically motivated, we do not 
advocate abandoning the definition based on origin.  Nevertheless, the separation 
into M dwarfs, ``brown dwarfs'', and giant ``planets'' is useful for parsing by eye the information in the figures.

In Figure 7, the bumps between 10$^{-4}$ \lo and 10$^{-3}$ \lo and between 10$^{8}$ and 10$^{9}$ years,
seen on the cooling curves of objects from 0.03 \mo to 0.08 \mo, are due to silicate and iron grain formation.   These effects,
first pointed out by Lunine \etal\ (1989), occur for T$_{\rm eff}$s between 2500 K and 1300 K.
The presence of grains affects the precise mass and luminosity at the edge of the main sequence.
Since grain and cloud models are problematic, there still remains much to learn concerning
their role and how to model them (\S 3.3 and Allard \etal\ 1997). 

Figure 8 depicts the central temperature (T$_c$) versus central density ($\rho_c$) 
for a variety of masses between 0.3\mj and 0.237 \mo.
Superposed are isochrones from 10$^{6.5}$ to 10$^{9.5}$ years.
For the M dwarfs, the central temperature generally rises until the object has stabilized as a star.  However, 
near the bottom edge of the main sequence the central temperature actually decreases slightly just before 
stabilizing.  The central density always increases with time.  The highest densities are achieved by
the massive brown dwarfs and hover near 1000 gm cm$^{-3}$ for solar metallicity.  They are higher for lower metallicities,
reaching a peak of $\sim$2000 gm cm$^{-3}$ at zero metallicity (Burrows \etal\ 1993).
The era during which T$_c$ increases is rather brief for the ``planets'' and they spend most of their time cooling
throughout.  The ``brown dwarfs'' nicely straddle these two regimes.

The Figure--8 trajectories for a given mass are universal curves, independent of the metallicity and 
atmosphere model.  They depend solely upon the equation of state and the fact that the structures are fully convective
in the interior.  However, the positions of the isochrones do depend upon the model specifics and vary
with metallicity and boundary conditions.  Note that the isochrones shear perceptibly near the 
``brown dwarf''--``planet'' interface.  Not unexpectedly, this is a consequence of the onset of deuterium burning.

For a given ``brown dwarf'' or EGP mass (in \mo), Figure 9 connects the 
observables T$_{\rm eff}$ (in K) and gravity ({\it g}, in $cgs$).  
The dashed curves are the isochrones.  As is clear from the figure,
gravity maps fairly directly onto mass and for no mass does g change by more than a factor of two 
after 10$^{8}$ years.  Modeling the spectrum of a substellar object will yield estimates of T$_{\rm eff}$ and g.
With these estimates, Figure 9 can be used to infer the mass and the age simultaneously.  
In fact, for a given composition and model, only two quantities are needed to derive {\it all} others.
Bolometric luminosity and age can be used to yield mass and radius.   Effective temperature
and mass can provide age and luminosity.  Our fit to the UKIRT spectrum of Gl229B (Marley \etal\ 1996; 
Geballe \etal\ 1996;  Matthews \etal\ 1996; see also Oppenheimer \etal\ 1995) 
gave T$_{\rm eff}$$\sim$900--1000 K and g$\sim$10$^{5\pm0.5}$ cm s$^{-2}$.
Reading off of Figure 9, one obtains a mass between 20 \mj and 60 \mj, with a best value near 35 \mj, and an age
between 10$^{8.5}$ and 10$^{9.5}$ years.  The wide range in inferred Gl229B parameters is a direct consequence 
of the weakness of our current constraints on g.

Figure 10 depicts the evolution of radius with T$_{\rm eff}$, which at later times is an erzatz age.
Jupiter's current radius is near $7\times10^9$ centimeters.
In some sense, a radius--T$_{\rm eff}$ plot
is a compact H--R diagram, since, while luminosities for the family range nine orders of magnitude
during a Hubble time, radii vary far less.
Note that initially it takes longer for more massive objects to shrink, but that isochrones are not very
much different from constant--radius lines at later times.  
Note also that the lowest mass objects ({\it e.g.}, Saturn)
tend to have larger radii at earlier times and smaller radii at later times.  In addition, at a given T$_{\rm eff}$,
the higher the mass the smaller the radius, while at a given age (after 10$^8$ years) radii generally 
decrease with mass.   Not obvious from the plot is the fact that the maximum cold radius occurs for a mass near
4$\!$ \mj.

Figure 11 is a theorist's H--R diagram for the ``brown dwarfs'' and giant ``planets.''
The inset is a continuation of the figure down to low luminosities and T$_{\rm eff}$s.
The current Jupiter and Saturn are superposed for comparison (Pearl \& Conrath 1991).  Importantly, constant mass
trajectories never cross and it is only for objects below 25$\!$ \mj that temperatures
below 400 K are reached within 10$^{10}$ years.  Figures 7 through 11 collectively summarize the
model space within which substellar objects reside.  Tables 1a-1d contain the results of evolutionary
calculations for objects with masses of 1 \mj, 5 \mj, 10 \mj, and 0.04 \mo ($\sim$42 \mj).
The numbers in them represent the latest atmospheric and opacity
calculations at solar metallicity.

\section{EGP and Brown Dwarf Spectra}

There are a few major aspects of EGP/brown dwarf atmospheres that bear listing and that uniquely
characterize them.  
Below T$_{\rm eff}$s of 1300 K, the dominant equilibrium carbon molecule is CH$_4$, not CO, 
and below 600 K the dominant nitrogen molecule is NH$_3$, not N$_2$.
As discussed in \S 2, the major opacity sources are H$_2$, H$_2$O, CH$_4$, and NH$_3$.
For T$_{\rm eff}$s below $\sim$400 K, water clouds form at or above the photosphere
and for T$_{\rm eff}$s below 200 K, ammonia clouds form ({\it viz.,} Jupiter).  Collision--induced absorption
of H$_2$ partially suppresses emissions longward of $\sim$10 \mic.  The holes in the opacity
spectrum of H$_2$O that define the classic telluric IR bands also regulate much of the emission from
EGP/brown dwarfs in the near infrared.  Importantly, the windows in H$_2$O and the suppression by H$_2$ conspire to
force flux to the blue for a given T$_{\rm eff}$.   
The upshot is an exotic spectrum enhanced relative to the blackbody value
in the $J$ and $H$ bands ($\sim$1.2 \mic\ and $\sim$1.6 \mic, respectively) by as much as {\it two} to {\it ten} orders of magnitude, 
depending upon T$_{\rm eff}$.  
In addition, as T$_{\rm eff}$ decreases below $\sim$1000 K, the flux in the $M$ band ($\sim$5 \mic)
is progressively enhanced relative to the blackbody value. 
While at 1000 K there is no enhancement, at 200 K it is near 10$^5$.   Hence, the $J$, $H$, and $M$ bands 
are the premier bands in which to search for cold substellar objects.   The $Z$ band ($\sim$1.05 \mic) is also 
super--blackbody over this T$_{\rm eff}$ range.  However, there is a NH$_3$ feature 
in the $Z$ band that is not in our database.  This
will likely reduce the flux in this band for the cooler models.   
Eventhough $K$ band ($\sim$2.2 \mic) fluxes are generally higher
than blackbody values,  H$_2$ and CH$_4$ absorption features in the $K$ band decrease its importance
{\it relative} to $J$ and $H$.  As a consequence of the increase of atmospheric 
pressure with decreasing T$_{\rm eff}$, the anomalously blue $J-K$ and $H-K$
colors get {\it bluer},
not redder.

For this paper, we calculated low--resolution spectra from 0.9 \mic\ to 2500 \mic, though we focus our discussions
on the 1 \mic\ to 10 \mic\ region.  A grid of spectra in T$_{\rm eff}$--gravity space was calculated.  As stated in \S 2.7,
for the evolutionary calculations described in
\S 4, a T$_{\rm eff}$--T$_{10}$ grid was also constructed. 
The evolutionary calculations were used to map mass and age onto T$_{\rm eff}$ and gravity, 
which were then used to interpolate in the grid of spectra to find the spectra and colors at 
any mass and age.  This procedure proved to be quite robust for T$_{\rm eff}$s between 1250 K and 100 K.
The spectra we present are for objects in isolation and ignore the transport effect of clouds.
As stated in \S 3.3, omitting the direct effects of clouds has consequences, for below 400 K the formation of
H$_2$O clouds partially depletes the spectrum of the H$_2$O vapor features that define it at higher tempertures.  
Note that the presence or absence of clouds strongly affects the reflection albedos of EGPs and brown
dwarfs.  In particular, when there are clouds at or above the photosphere, the albedo in the optical
is high.  Conversely, when clouds are absent, the albedo in the mostly absorbing atmosphere
is low.

Figure 12 depicts the object's surface flux versus wavelength for representative T$_{\rm eff}$s from 130 K to
1000 K at a gravity of $3.0\times10^4$ cm s$^{-2}$.  The corresponding masses range 
from $\sim$13 \mj to $\sim$16 \mj and the corresponding ages range from 0.25 Gyrs to 7 Gyrs.  
Superposed on Figure 12 are the positions of various prominent molecular bands and the $J$, $H$,
$K$, and $M$ bands.  As is clear from the figure, H$_2$O defines much of the spectrum, but CH$_4$ and 
H$_2$ modify it in useful ways.  CH$_4$ absorption features near 1.65 \mic, 2.2 \mic, and 3.3 \mic\
are particularly relevant, the latter two decreasing the $K$ and $L^{\prime}$ ($\sim$3.5 \mic) band fluxes, respectively.  
NH$_3$ near 6 \mic\ becomes important below 250 K and the CH$_4$
feature around 7.8 \mic\ deepens with decreasing  T$_{\rm eff}$.   However, it should be noted
that in Jupiter the 7.8 \mic\ absorption feature is inverted into a stratospheric emission feature.
Since a stratosphere requires UV flux from the primary or another energy deposition mechanism, 
our models do not address this possibility.
In addition, we find that H$_2$O and NH$_3$ features near 6 \mic\ make the band from 5.5 to 7 \mic\
less useful for searching for brown dwarfs and EGPs.   

Figure 13 depicts spectra between 1 \mic\ and 40 \mic\ at a detector 10 parsecs away from objects with age 1 Gyr and masses
from 1 \mj through 40 \mj.  Figure 14 portrays the spectra for the same parameters, but from 1 \mic\ 
to 10 \mic.  Superposed on the former are the corresponding blackbody curves and superposed on both are putative sensitivities
for the three NICMOS cameras (Thompson 1992), ISO (Benvenuti \etal\ 1994), SIRTF (Erickson \& Werner 1992), 
and Gemini/SOFIA (Mountain \etal\ 1992; Erickson 1992).  
Figure 13 demonstrates how unlike a blackbody an EGP spectrum
is.  Note on Figure 13 the H$_2$--induced suppression at long wavelengths and the enhancement at shorter
wavelengths.  For example, the enhancement at 5 \mic\ for a 1 Gyr old, 1 \mj$\!$ extrasolar planet is by four orders
of magnitude.  Implicit in Figure 13 is the enhancement around the N band ($\sim$10 \mic)
for T$_{\rm eff}$ below 200 K.

Comparison with the sensitivities reveals that the range for detection by SIRTF at 5 \mic\ 
of a 1 Gyr old, 1 \mj object in isolation is near 100 parsecs.   The range for NICMOS in $H$
for a 1 Gyr old, 5 \mj object is approximately 300 parsecs, while for a coeval 40 \mj object
it is near 1000 parsecs.  These are dramatic numbers that serve to illustrate both the
promise of the new detectors and the enhancements we theoretically predict.

Figures 15--19 portray the evolution from 0.1 Gyr to 5 Gyr of the spectra from 1 \mic\ to 10 \mic\ 
of objects with masses of 1, 5, 10, 15, 20 \mj.  The higher curves are for the younger ages.
These cooling curves summarize EGP/brown dwarf 
spectra and their evolution, but are merely representative of the suite of models now available.
Note that the scales change from Figures 15--16 to Figures 17--19 and that, for comparison, blackbody curves are superposed
on Figure 15 (the ``Jupiter'' model).
Figures 17 and 19 include identifications of some of the molecular features.
Figure 19 suggests that SIRTF will be able to see at 5 \mic\ a 5 Gyr old, 20 \mj object in isolation out to 
$\sim$400 parsecs and that NICMOS will be able to see at $J$ or $H$ a 0.1 Gyr old object with the same
mass out to $\sim$2000 parsecs.  As shown in Figure 15, the $J$ and $H$ flux enhancements over blackbody values
for the 1 \mj$\!$ model after 0.1 Gyr are at least ten orders of magnitude.   However, it must be remembered that
these models do not include a reflected light component from a primary.  For many combinations
of primary and orbital separation, this reflected component can dominate in the near IR.

\section{Infrared Colors}

From the spectra described in the previous section, we have calculated infrared colors and produced 
color--color and color--magnitude diagrams.    Figures 20 through 24 are representative color--magnitude
diagrams for objects with masses from 3 \mj to 40 \mj, for ages of 0.5, 1.0, and 5.0 Gyr.  
Figures 25 and 26 are color--color diagrams for the same models.   
For comparison, included in these figures are the corresponding blackbody curves, 
hot, young brown dwarf or extremely late M dwarf
candidates such
as LHS2924, GD 165B, Calar 3, and Teide 1 (Kirkpatrick, Henry, \& Simons 1994,1995; Zapatero-Osorio, Rebolo, \& Martin 1997),
and a sample of M dwarfs from Leggett (1992).  
These figures collectively illustrate the unique color realms occupied by extrasolar giant planets and brown dwarfs.

Figures 20 and 21 portray the fact that the $K$ and $J$ versus $J-K$ infrared H--R diagrams loop back to the blue
below the edge
of the main sequence and are not continuations of the M dwarf sequence into the red.
The difference between the blackbody curves and the model curves is between 3 and 10 magnitudes
for $J$ versus $J-K$, more for $K$ versus $J-K$.  Gl229B fits nicely on these theoretical isochrones.
The suppression of $K$ by H$_2$ and CH$_4$ features is largely responsible for this anomalous blueward
trend with decreasing mass and T$_{\rm eff}$.
As Figures 22 and 23 demonstrate, the fit to Gl229B in $H$ is not as good.
This is also true of the fit to $L^{\prime}$.   Since both $H$ and  $L^{\prime}$ have
significant CH$_4$ features in them, we surmise that incompleteness 
or errors in the CH$_4$ opacity database is the culprit.
As Figures 22 and 23 also show, $J-H$ actually reddens with decreasing T$_{\rm eff}$, but only marginally and
is still 1.5 to 4 magnitudes bluer than the corresponding blackbody.
That the $J-H$ and $H-K$ colors of EGPs and brown dwarfs
are many magnitudes blueward of blackbodies is a firm conclusion of this work. 

Superposed on the color--color diagrams (Figures 25 and 26) are model colors for stars at the edge of the main sequence
for metallicities from solar to 10$^{-3}$ times solar.  For the non--solar calculations,  
the atmospheres of Allard \& Hauschildt
(1995) were used to generate the corresponding T$_{\rm eff}$--T$_{10}$ relations
employed by our evolutionary code (Saumon \etal , in preparation).  
A glance at these numbers and those in the zero--metallicity paper
of Saumon \etal\ (1994) reveals that we expect the lower metallicity models to populate the bluer
regions below the depicted model lines.  However, we have yet to calculate precise numbers for
non--solar metallicities with the new algorithms and opacities of this paper.

Tables 2 and 3 depict the infrared magnitudes and colors for various gravities and T$_{\rm eff}$s.
Also included are $N$ band magnitudes and $M-N$ colors.  
We employed the transmission curves of Bessell \& Brett \markcite{bb88}(1988) and 
Bessell \markcite{bessell90}(1990) to define the photometric bandpasses 
and the model of Vega by Dreiling \& Bell \markcite{db80}(1980) for the calibration
of the magnitude scale.  As Table 2 and Figures 20--24 suggest, the brightnesses
in the near IR are quite respectable. Table 3 shows that colors generally get bluer with increasing
gravity (except for $K-L^{\prime}$, which shows the opposite trend).  
For $J-H$, the effect may be only $\sim$0.2 magnitudes per
decade in gravity and for the problematic $H-K$ color it is perhaps $\sim$0.4 magnitudes per decade.  However,
for $K-L^{\prime}$ it is $\sim$0.8 magnitudes per decade, 
though one must recall that $L^{\prime}$ is not well modeled.
Nevertheless, in principle these colors can collectively be used as crude gravity diagnostics.

\section{Conclusions and Future Work}

During the past two years, scientists and the public at large have been
galvanized by the discovery of planets and brown dwarfs around nearby stars and by evidence
for ancient life on Mars (McKay \etal\ 1996).
These extraordinary findings have dramatically heightened
interest in the age--old questions of where we came from and whether we are unique in the cosmos.
NASA has outlined a program to detect planetary
systems around nearby stars that may become a future focus of NASA
and its central, unifying scientific theme in the next century.
This vision is laid out in the {\it Exploration of Neighboring Planetary Systems (ExNPS)
Roadmap} (see also the ``TOPS'' Report, 1992) and has been expanded to include the Origins
of life, planets, galaxies, and the universe.

The next generation planet and brown dwarf searches and studies will be conducted by
NICMOS, SIRTF, Gemini/SOFIA,
ISO, NGST, LBT (Angel 1994), the MMT conversion, the VLT, Keck I \& II, COROT (transits), DENIS, 2MASS,
UKIRT, and IRTF, among other platforms.   For close companions,
advances in adaptive optics, interferometry, and coronagraphs
will be necessary to disentangle the light of companion and primary.

The models we have generated of the colors and spectra of EGPs and brown
dwarfs are in aid of this quest for Origins and of the discovery and characterization of 
substellar objects around nearby stars and in the field.  
We have created a general non--gray theory of objects from 0.3 \mj to 70 \mj below $\sim$1300 K using the best
input physics and some of the best numerical tools available, but much remains to be done.  In particular,
the opacity of CH$_4$ and a proper treatment of silicate/iron, H$_2$O, and NH$_3$ clouds are future challenges
that must be met before the theory can be considered mature. 
Furthermore, the effects of stellar insolation, addressed only approximately 
in Saumon \etal\ (1996) and Guillot \etal\ (1996),
must be incorporated consistently. 
Since the near IR signature of proximate substellar companions will be
significantly altered by a refected component, a theory of albedos in the optical and in the near IR must be developed. 
For specificity, we focussed in this paper upon objects in isolation and did not 
include the complicating parameters of central star and semi--major axis. 
However, it will be useful to predict the signatures
of specific systems with known orbital characteristics, primaries, and ages,
such as $\tau$ Boo, 51 Peg, $\upsilon$ And,
55 Cnc, $\rho$ CrB, 70 Vir, 16 Cyg, and 47 UMa.

Nevertheless, our theoretical calculations lead to certain general conclusions:

\begin{enumerate}

        \item H$_2$O, H$_2$, and CH$_4$ dominate the spectrum below T$_{\rm eff}$$\sim$1200 K.
For such T$_{\rm eff}$s, most or all true metals are sequestered below the photosphere.
        \item Though EGP colors and low--resolution spectra depend upon gravity, 
this dependence is weak. However, high--resolution
spectra may provide useful gravity diagnostics. 
        \item The primary bands in which to search are $Z$, $J$, $H$, $K$, $M$ and $N$.
$K$ is not as good as $J$ or $H$.
        \item Enhancements and suppressions of the emergent flux relative to blackbody values can be by
many orders of magnitude.
        \item Objects that were considered from their low T$_{\rm eff}$s ($\sles\ $600 K) to be undetectable
in the near IR may not be.  
        \item The infrared colors of EGPs and brown dwarfs are much bluer than the colors previously
derived using either the blackbody assumption or primitive non--gray models. 
        \item In some IR colors ({\it e.g.,} $J-K$), an object gets bluer, not redder, with age
and for a given age, lower--mass substellar objects are bluer than higher--mass substellar objects.
        \item For a given composition, only two observables are necessary to 
constrain a substellar object's parameters.  For instance, given only T$_{\rm eff}$ and
gravity, one can derive mass, age, and radius.
        \item The existence of an interior radiative zone seems to be a generic feature of
substellar objects with T$_{\rm eff}$s from $\sim$200 K to 1000 K, and might also obtain 
for T$_{\rm eff}$s below $\sim$200 K.  The appearance and extent of such a radiative zone is
a function of gravity.
        \item Clouds of H$_2$O and NH$_3$ are formed for T$_{\rm eff}$s 
below $\sim$400 K and $\sim$200 K, respectively.  Their formation will affect the colors and spectra
of EGPs and brown dwarfs in ways not yet fully characterized.

\end{enumerate}

\acknowledgements 

We thank R. Angel, W. Benz, S. Kulkarni, J. Liebert, B. Oppenheimer, G. Rieke, G. Schneider, S. Stolovy,
and N. Woolf for a variety of useful contributions.  
This work was supported under NSF grants AST-9318970 and AST-9624878 and under
NASA grants NAG5-2817, NAGW-2250, and NAG2-6007. 



\newpage

\figcaption{
T$_{\rm eff}$--T$_{10}$--$g$ surface used for
evolutionary calculations presented in this paper.
}

\figcaption{
Atmospheric pressure--temperature profiles
for EGPs with surface gravity fixed at 2200 cm s$^{-2}$ and T$_{\rm eff}$ =
1000, 900, 800, 700, 600, 500, 400, 300, 200, and 128 K.
}

\figcaption{
Atmospheric pressure--temperature profiles
for EGPs with surface gravity fixed at $10^4$ cm s$^{-2}$ and T$_{\rm eff}$ =
800, 600, 500, 400, 200, and 128 K.
}

\figcaption{
Atmospheric pressure--temperature profiles
for EGPs with surface gravity fixed at $3 \times 10^4$ cm s$^{-2}$
and T$_{\rm eff}$ =
1100, 900, 700, 300, 200, and 128 K.
}

\figcaption{
Atmospheric pressure--temperature profiles
for EGPs with surface gravity fixed at $10^5$ cm s$^{-2}$ and T$_{\rm eff}$ =
1200, 1100, 900, 700, 500, 250, 200, and 128 K.
}

\figcaption{
Atmospheric pressure--temperature profiles
for EGPs with surface gravity fixed at $3 \times 10^5$ cm s$^{-2}$
and T$_{\rm eff}$ =
1200, 1100, 900, 700, 200, and 128 K.
}

\figcaption{
Evolution of the luminosity (in L${_\odot}$) of solar--metallicity M dwarfs and substellar objects
versus time (in years) after formation.
The stars, ``brown dwarfs'' and ``planets'' are shown as solid, dashed, and dot--dashed
curves, respectively.  
In this figure, we arbitrarily designate as ``brown dwarfs'' those objects that burn deuterium,
while we designate those that do not as ``planets.''
The masses in M${_\odot}$ label most of the curves, with the lowest three
corresponding to the mass of Saturn, half the mass of Jupiter, and the mass of Jupiter.
}

\figcaption{
Evolutionary tracks of central density (in gm cm$^{-3}$) versus 
central temperature (in K) for stars (solid), ``brown dwarfs'' (dashed) and ``giant
planets''(dot--dashed), as in Figure 7.  The isochrones
are drawn as gray curves and are labeled in log$_{10}$ years.
The pronounced wave in the isochrones between about $log_{10} T_c$ = 5.5 and 6 is due to
deuterium burning.
A given mass defines a unique relationship between
central temperature and density which is independent of metallicity. The only effect
of the metallicity is to change the rate at which the central temperature and density
evolve and the positions of the isochrones.
}

\figcaption{
Evolutionary tracks of log$_{10}$ gravity (in cm s$^{-2}$) versus effective
temperature (in K) for ``brown dwarfs'' (solid) and ``planets'' (dashed).
The isochrones are gray curves and are labeled in log$_{10}$ years.
In all cases, gravity increases with time.  Initially for the more
massive brown dwarfs, the effective temperature is roughly constant, or
slightly increasing, before decreasing inexorably at later times.
This figure depicts how T$_{\rm eff}$ and gravity map onto mass and age.
}

\figcaption{
log$_{10}$ radius (in centimeters) versus effective temperature (T$_{\rm eff}$, in K), with T$_{\rm eff}$
decreasing to the right. This plot has the advantage over an H-R diagram that considerably more
detail can be shown over the range of conditions considered.
In all cases, radius decreases with time. As depicted in Figure 9, for the more massive brown dwarfs
the effective temperature initially increases before decreasing. 
}

\figcaption{
H-R diagram: luminosity (in \lo) versus T$_{\rm eff}$ (in K) for various masses
labeled on the figure in \mo.  Due to the large range in
luminosity and the near degeneracy of the tracks of substellar objects at late stages of
evolution, it is not possible to represent with adequate detail the whole H-R diagram as
one figure.  Accordingly, the low--temperature and low--luminosity tail of the H-R diagram is
shown in the inset; note that the axes are scaled differently, but otherwise correspond to
those on the main figure.  For additional clarity, several masses have been omitted in the inset.
We have labeled the observed positions of Jupiter and
Saturn as points ``J'' and ``S,'' respectively (Pearl \& Conrath 1991).
As discussed in Figure 7, all substellar objects decrease in luminosity monatonically,
though during the early phases deuterium burning slows the evolution.  As the ``brown
dwarfs'' and ``planets'' cool to their cold radii, their tracks in the lower right of the H-R
diagram correspond closely to curves of constant radius.  Moreover, in the late phases of evolution, due to the very
weak dependence of radius on mass, the curves of the lower--mass objects become degenerate.
}

\figcaption{
Surface flux (in erg cm$^{-2}$ sec$^{-1}$ Hz$^{-1}$) versus wavelength (in microns) 
from 1 \mic\ to 10 \mic\ for T$_{\rm eff}$s of 130, 200, 300, 500, 600, 700, and 1000 K,
at a surface gravity of $3.0\times10^4$ cm s$^{-2}$.  Shown are the positions of the $J$,
$H$, $K$, and $M$ bands and various molecular absorption features.  See the text for discussion.
}

\figcaption{
The flux (in $\mu$Janskys) at 10 parsecs versus wavelength (in microns) from 1 \mic\ to
40 \mic\ for 1, 5, 10, 20, 30, and 40\mj models at 1 Gyr.   Superposed for comparison
are the corresponding blackbody curves (dashed) and the putative sensitivities of the three NICMOS cameras, ISO,
Gemini/SOFIA, and SIRTF.  NICMOS is denote with large black dots, ISO with thin, dark lines,
Gemini/SOFIA with thin, light lines, and SIRTF with thicker, dark lines. 
At all wavelengths, SIRTF's projected sensitivity is greater than ISO's.
SOFIA's sensitivity overlaps with that of ISO around 10\mic.  For other wavelength intervals, the order
of sensitivity is SIRTF $>$ Gemini/SOFIA $>$ ISO, where $>$ means ``is more sensitive than.'' 
Note the suppression relative to the blackbody values at the longer
wavelengths.
}
\figcaption{
The flux at 10 parsecs (in $\mu$Janskys) versus wavelength (in microns) for the same models
depicted in Figure 13, but for a wavelength range of 1 \mic\ to 10 \mic.  Shown are the positions of the $J$,
$H$, $K$, and $M$ bands and various molecular absorption features.  Also included are the estimated 
sensitivities of NICMOS, ISO, Gemini/SOFIA, and SIRTF, as described in the caption to Figure 13.
}

\figcaption{
The flux (in $\mu$Janskys) at 10 parsecs versus wavelength (in microns) from 1 \mic\ to
10 \mic\ for a 1$\!$ \mj object at ages of 0.1, 0.5, 1.0, and 5.0 Gyr.
Superposed are the positions of the $J$, $H$, $K$, and $M$ bands and the corresponding blackbody
curves (dashed), as well as the estimated sensitivities of the three NICMOS cameras, ISO, Gemini/SOFIA,
and SIRTF (see caption to Figure 13).
}

\figcaption{
The flux (in $\mu$Janskys) at 10 parsecs versus wavelength (in microns) from 1 \mic\ to
10 \mic\ for a 5 \mj object at ages of 0.1, 0.5, 1.0, and 5.0 Gyr.
Superposed are the positions of the $J$, $H$, $K$, and $M$ bands 
and the estimated sensitivities of the three NICMOS cameras, ISO, Gemini/SOFIA,
and SIRTF.
}

\figcaption{
The flux (in $\mu$Janskys) at 10 parsecs versus wavelength (in microns) from 1 \mic\ to
10 \mic\ for a 10 \mj object at ages of 0.1, 0.5, 1.0, and 5.0 Gyr.
Superposed are the positions of the $J$, $H$, $K$, and $M$ bands, 
the estimated sensitivities of the three NICMOS cameras, ISO, Gemini/SOFIA,
and SIRTF, and the positions of various of the important molecular absorption features.
}

\figcaption{
The flux (in $\mu$Janskys) at 10 parsecs versus wavelength (in microns) from 1 \mic\ to
10 \mic\ for a 15 \mj object at ages of 0.1, 0.5, 1.0, and 5.0 Gyr.
Superposed are the positions of the $J$, $H$, $K$, and $M$ bands 
and the estimated sensitivities of the three NICMOS cameras, ISO, Gemini/SOFIA,
and SIRTF.
}

\figcaption{
The flux (in $\mu$Janskys) at 10 parsecs versus wavelength (in microns) from 1 \mic\ to
10 \mic\ for a 20 \mj object at ages of 0.1, 0.5, 1.0, and 5.0 Gyr.
Superposed are the positions of the $J$, $H$, $K$, and $M$ bands,
the estimated sensitivities of the three NICMOS cameras, ISO, Gemini/SOFIA,
and SIRTF, and the positions of various of the important molecular absorption features.
}

\figcaption{
Absolute $J$ vs. $J-K$ color--magnitude diagram.  Theoretical isochrones
are shown for $t$ = 0.5, 1, and 5 Gyr, along with their blackbody
counterparts.  The difference between blackbody colors and model
colors is striking.  The brown dwarf, Gliese 229B (Oppenheimer \etal\ 1995),
the young brown dwarf candidates Calar 3 and Teide 1 (Zapatero-Osorio, Rebolo, \& Martin 1997), 
and late M dwarfs LHS 2924 and 
GD165B (Kirkpatrick, Henry, \& Simons 1994,1995)) are plotted for comparison.
The lower main sequence is defined by a selection of M--dwarf stars from 
Leggett (1992).
}

\figcaption{
Absolute $K$ vs. $J-K$ color--magnitude diagram.  Otherwise as in Figure 20.
}

\figcaption{
Absolute $J$ vs. $J-H$ color--magnitude diagram.  Otherwise as in Figure 20.
}

\figcaption{
Absolute $H$ vs. $J-H$ color--magnitude diagram.  Otherwise as in Figure 20.
}

\figcaption{
Absolute $H$ vs. $H-K$ color--magnitude diagram.  Otherwise as in Figure 20.
}

\figcaption{
$J-H$ vs. $H-K$ color--color diagram.  
The edge of the main sequence as a function of metallicity, from our calculations employing Allard \& Hauschildt (1995)
atmosphere models, is shown for metallicities from 
[M/H]=0 (top) to [M/H]=-3 (bottom) (Saumon \etal\, in preparation).
Otherwise as in Figure 20.
}

\figcaption{
$J-K$ vs. $K-L^{\prime}$ color--color diagram.  Otherwise as in Figure 25.
}


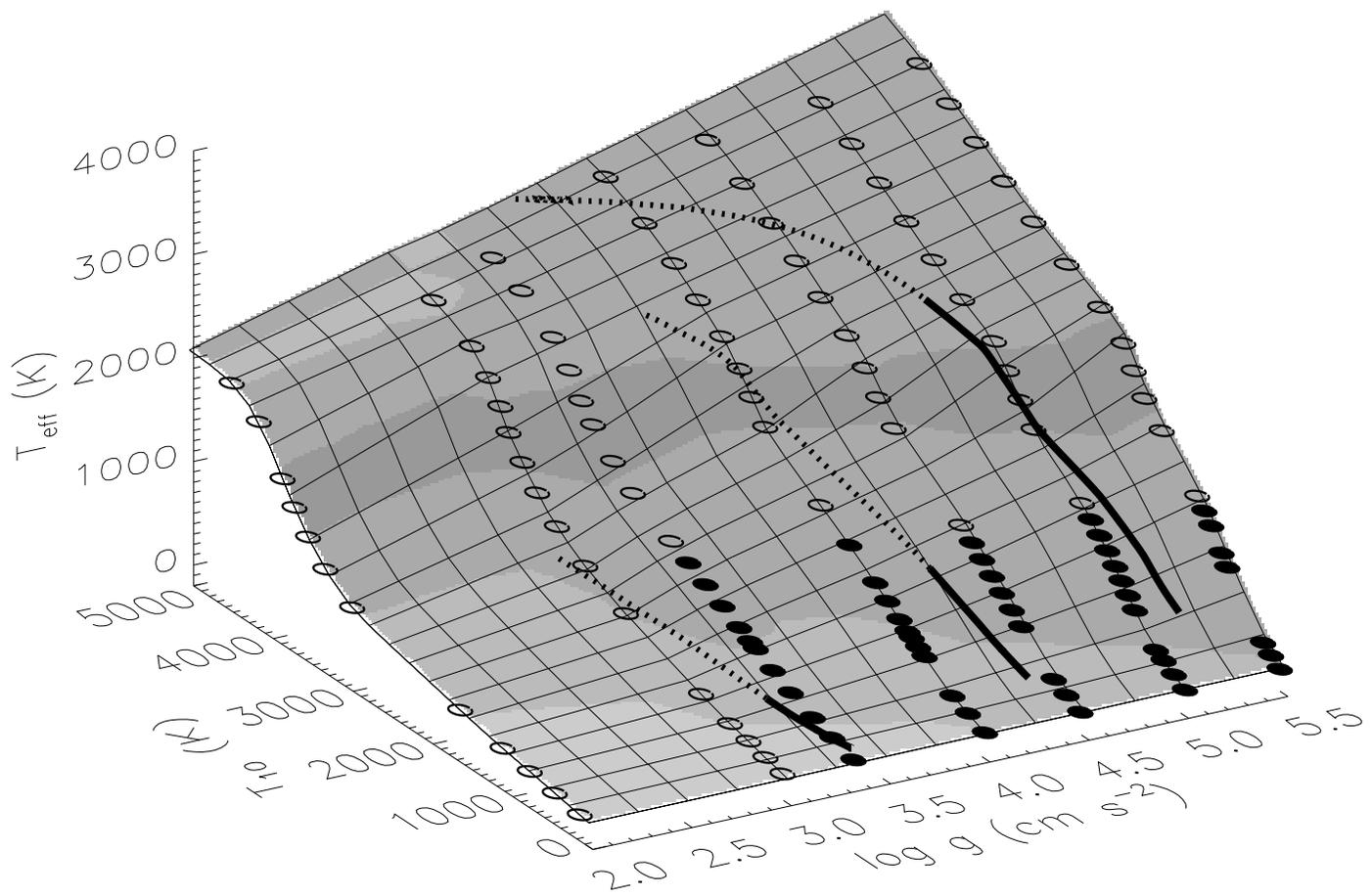
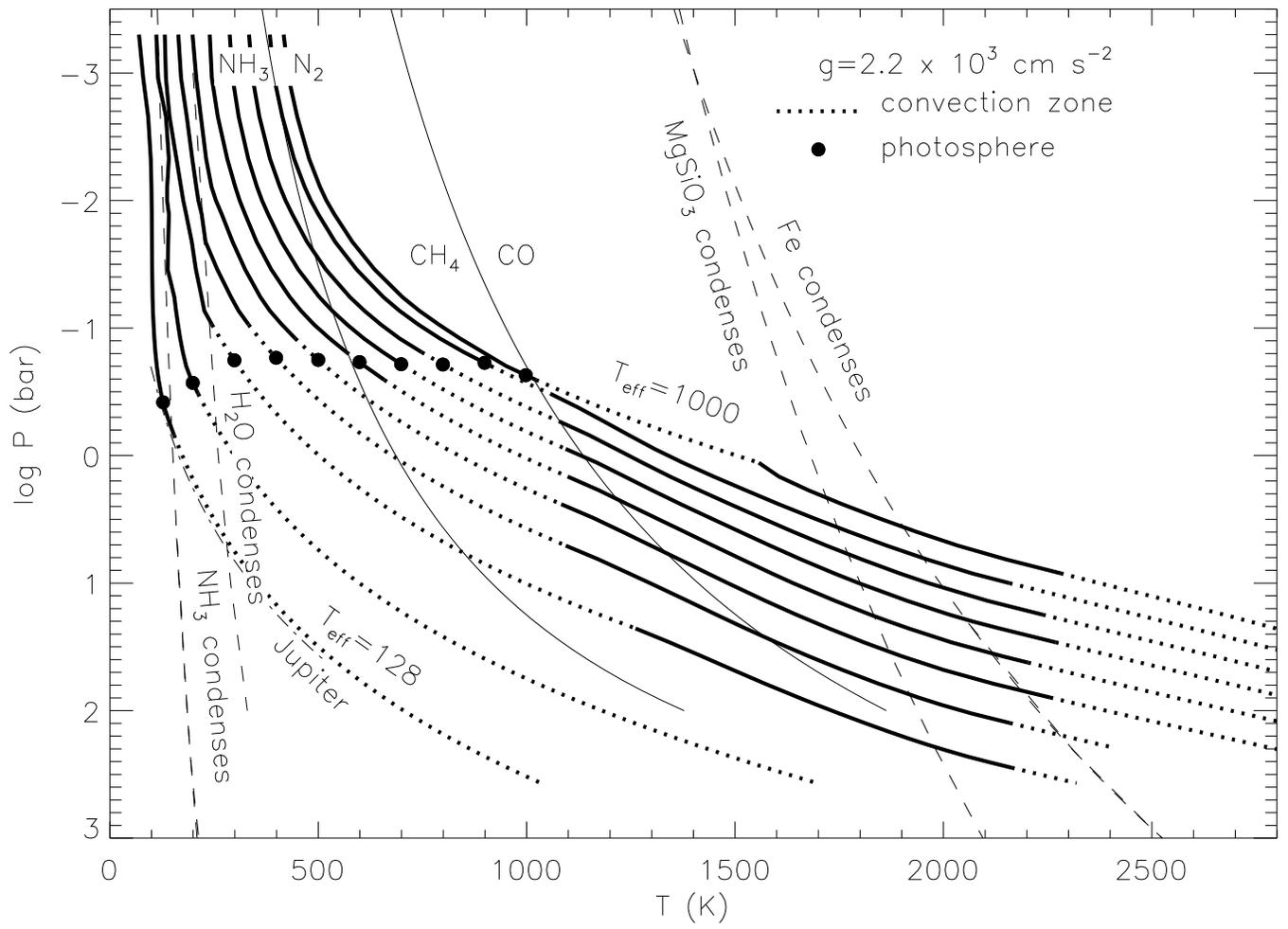
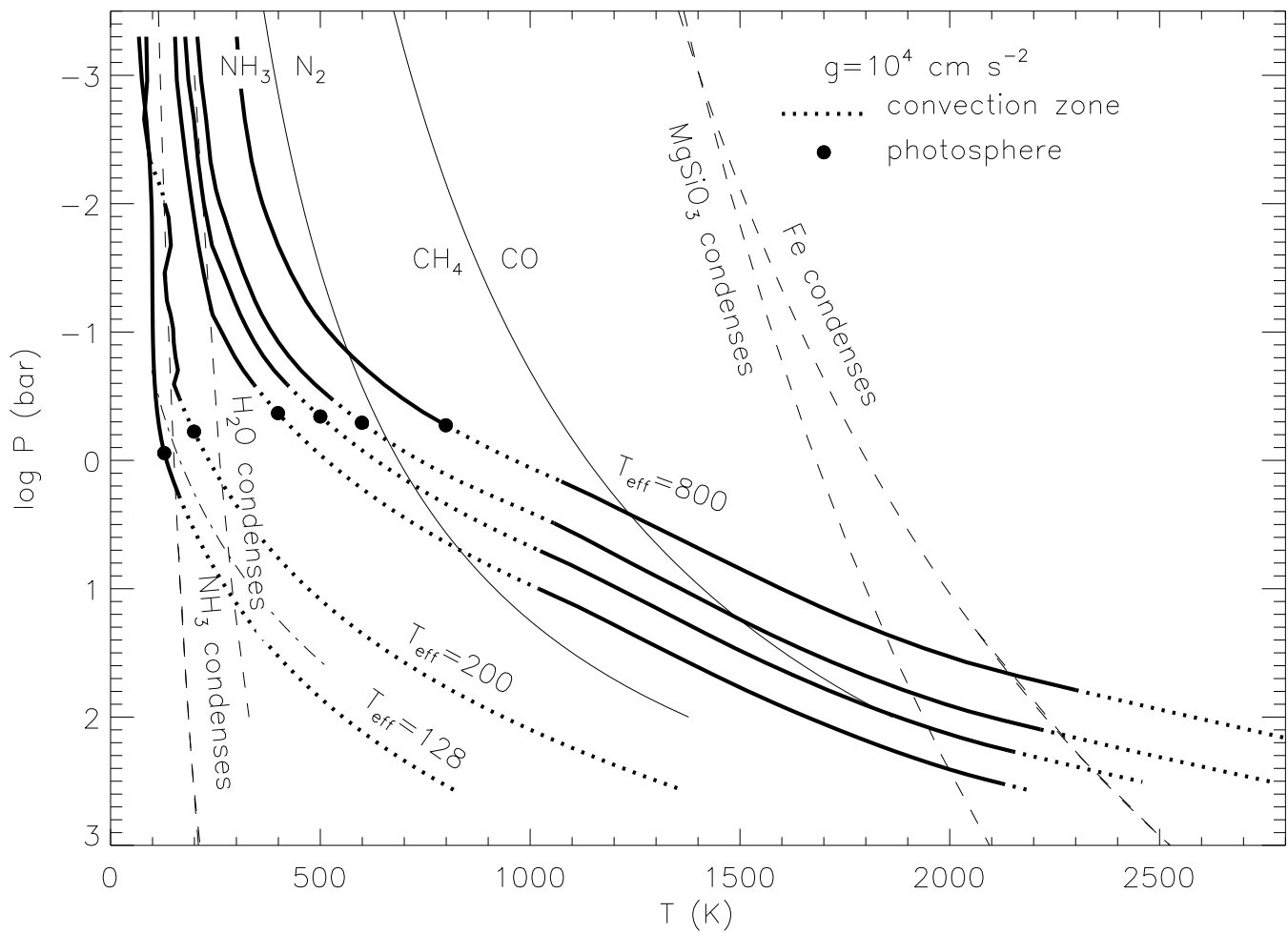
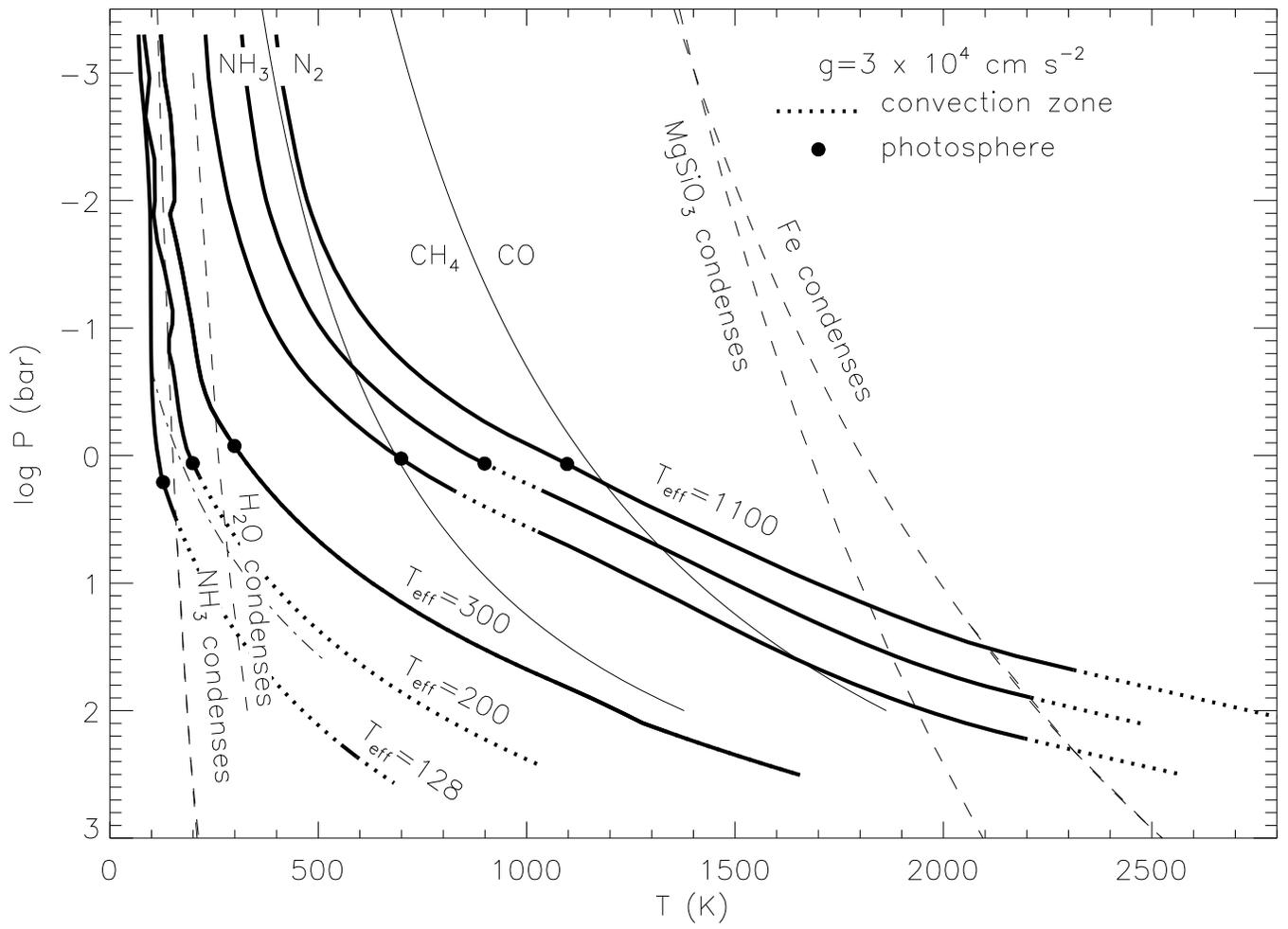
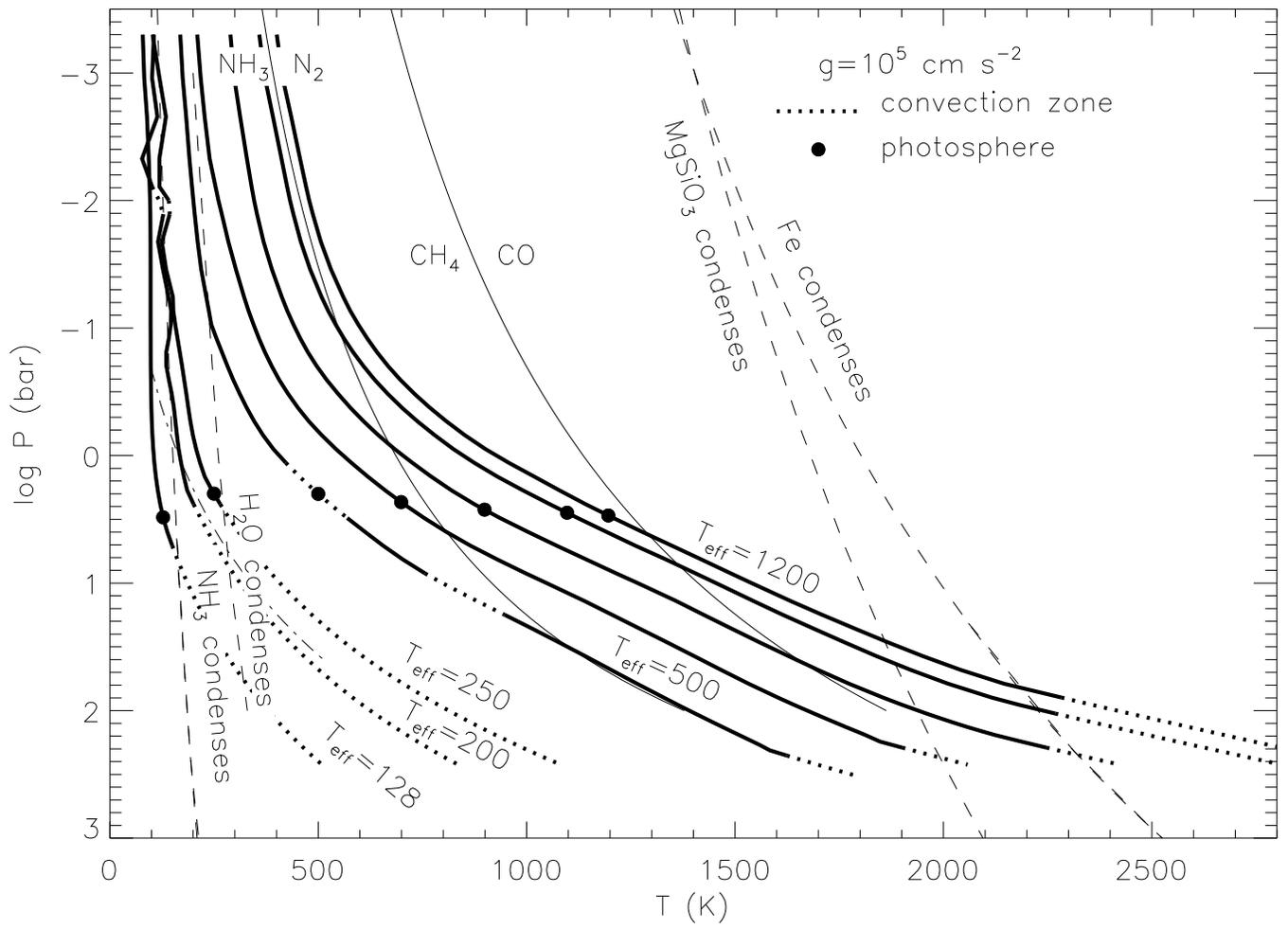
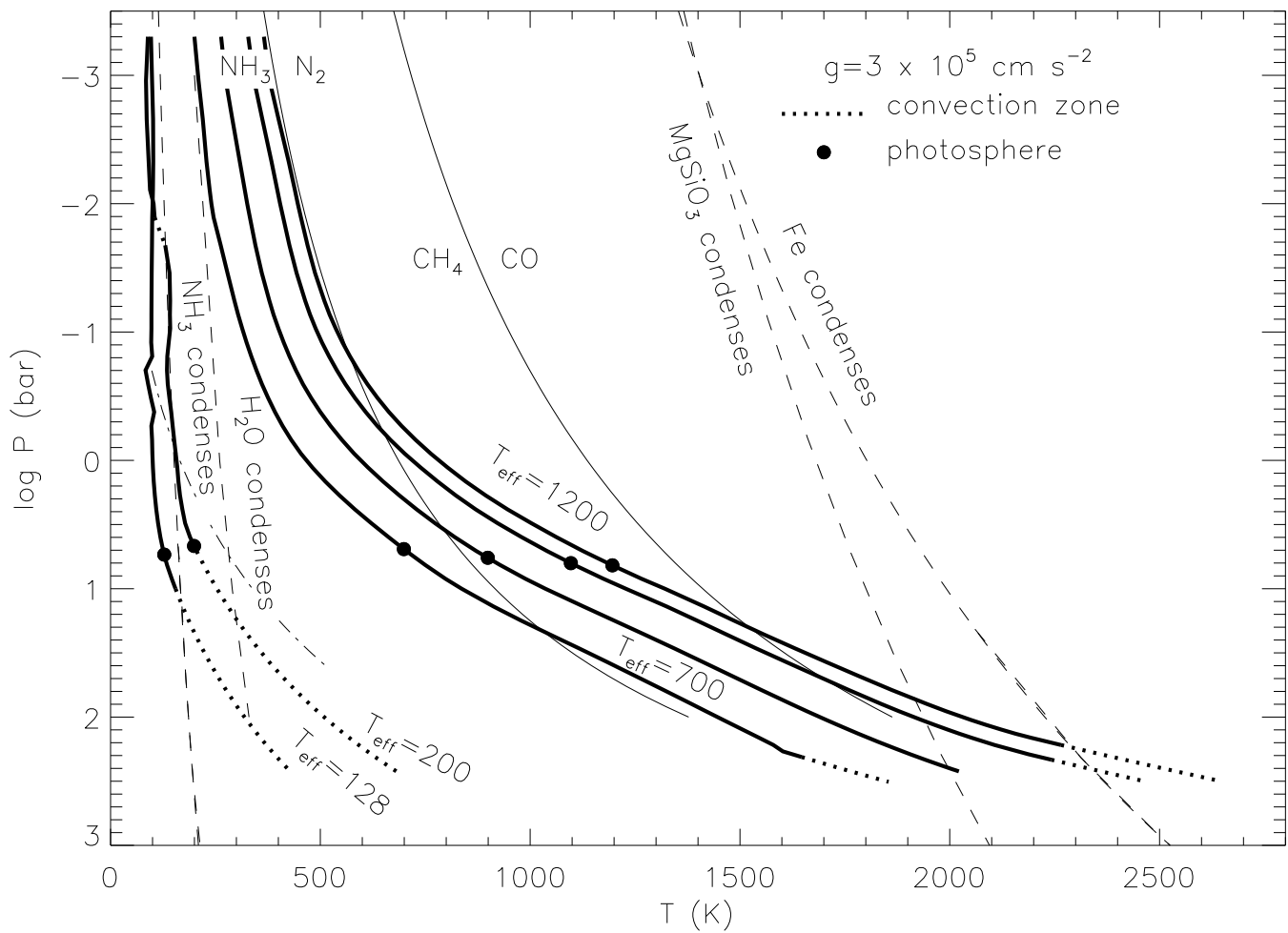
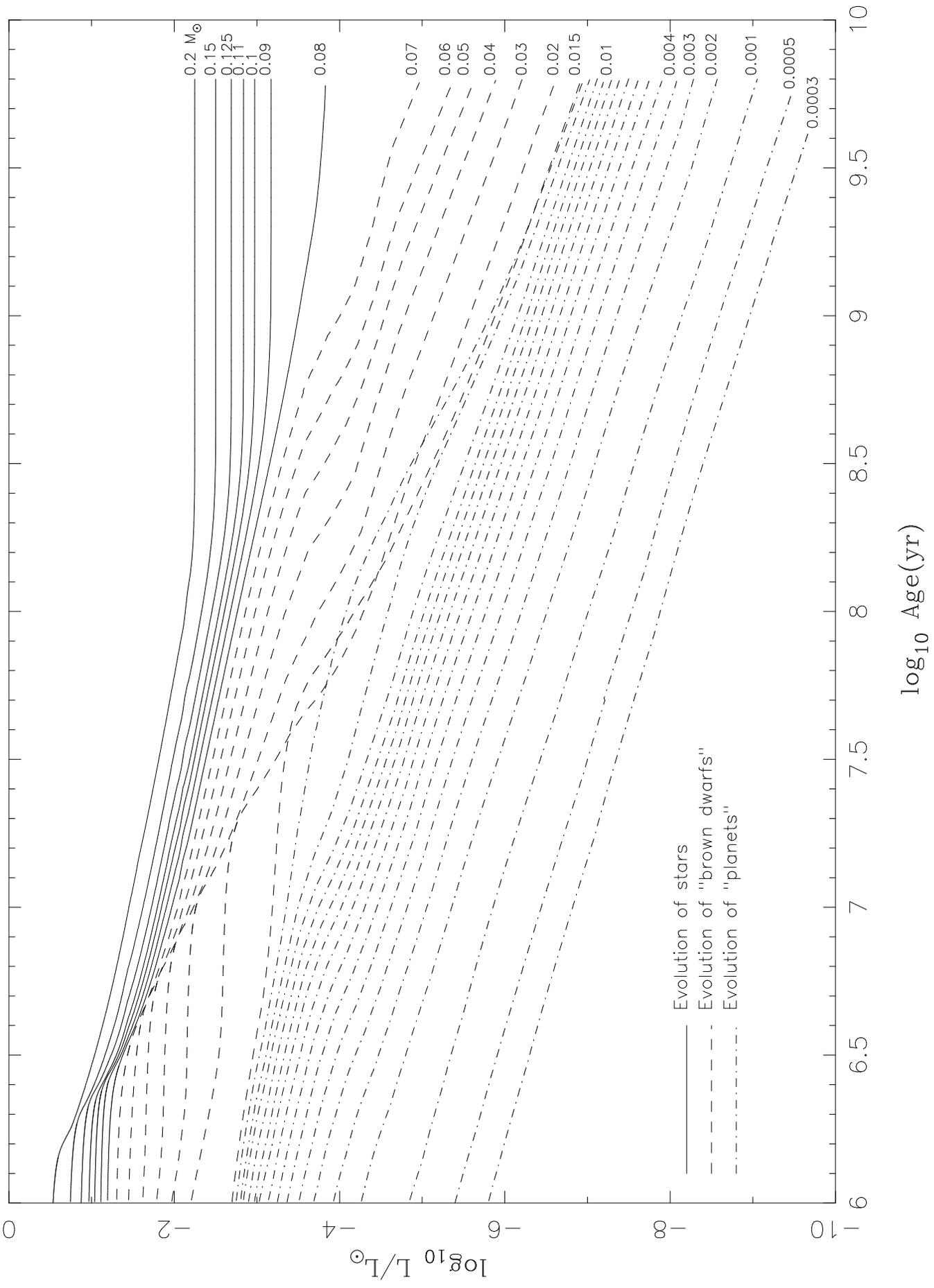
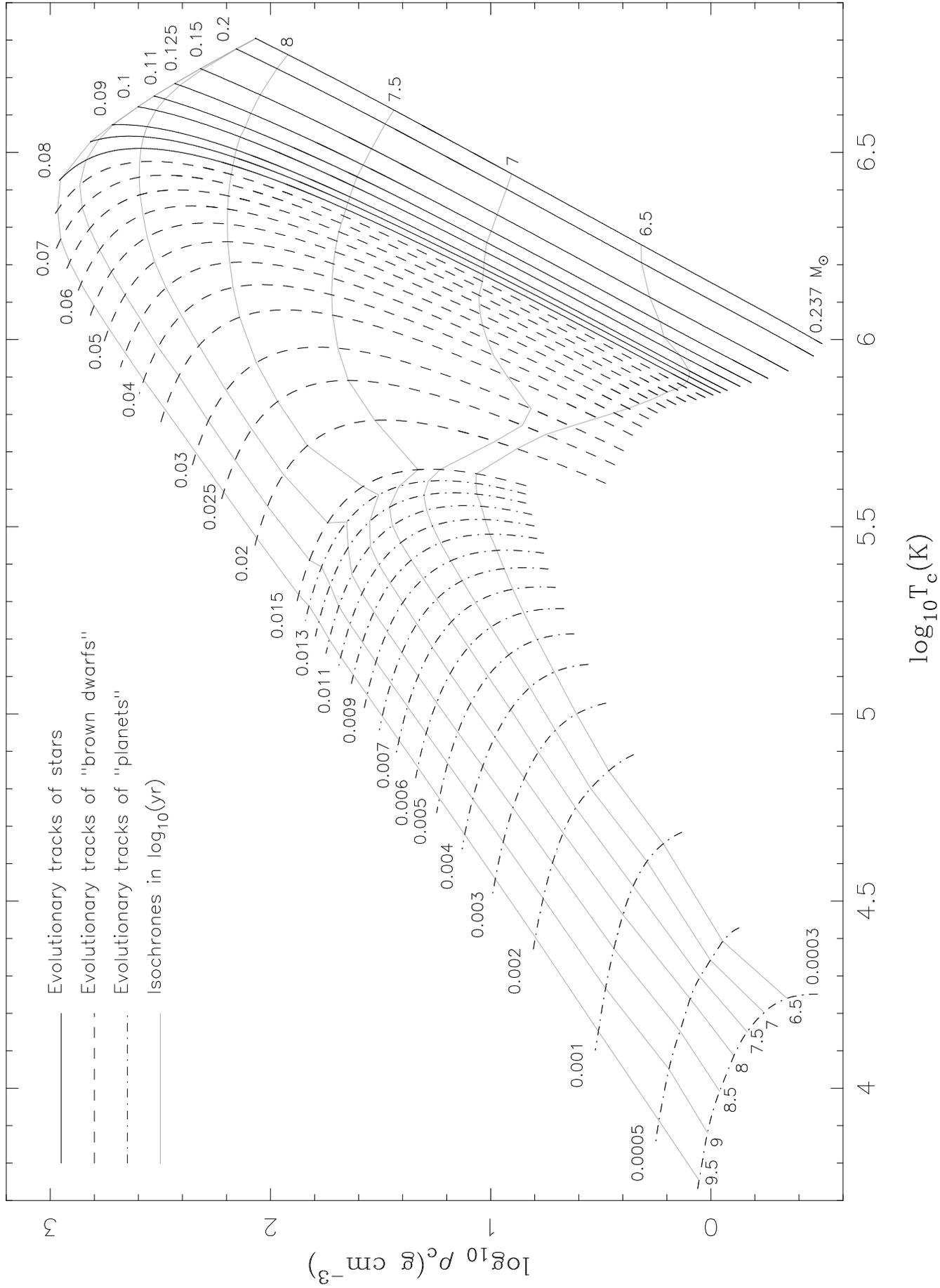
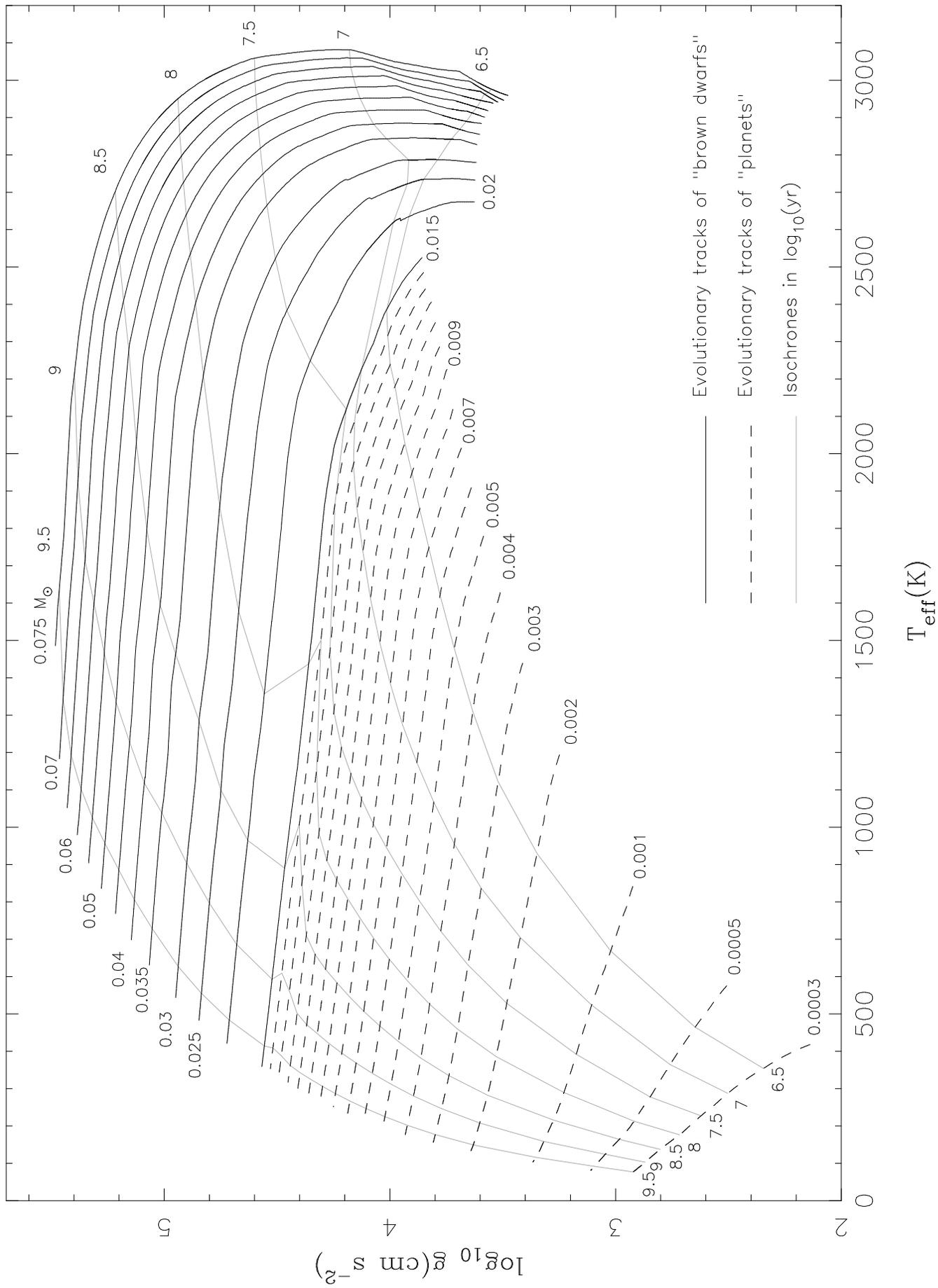
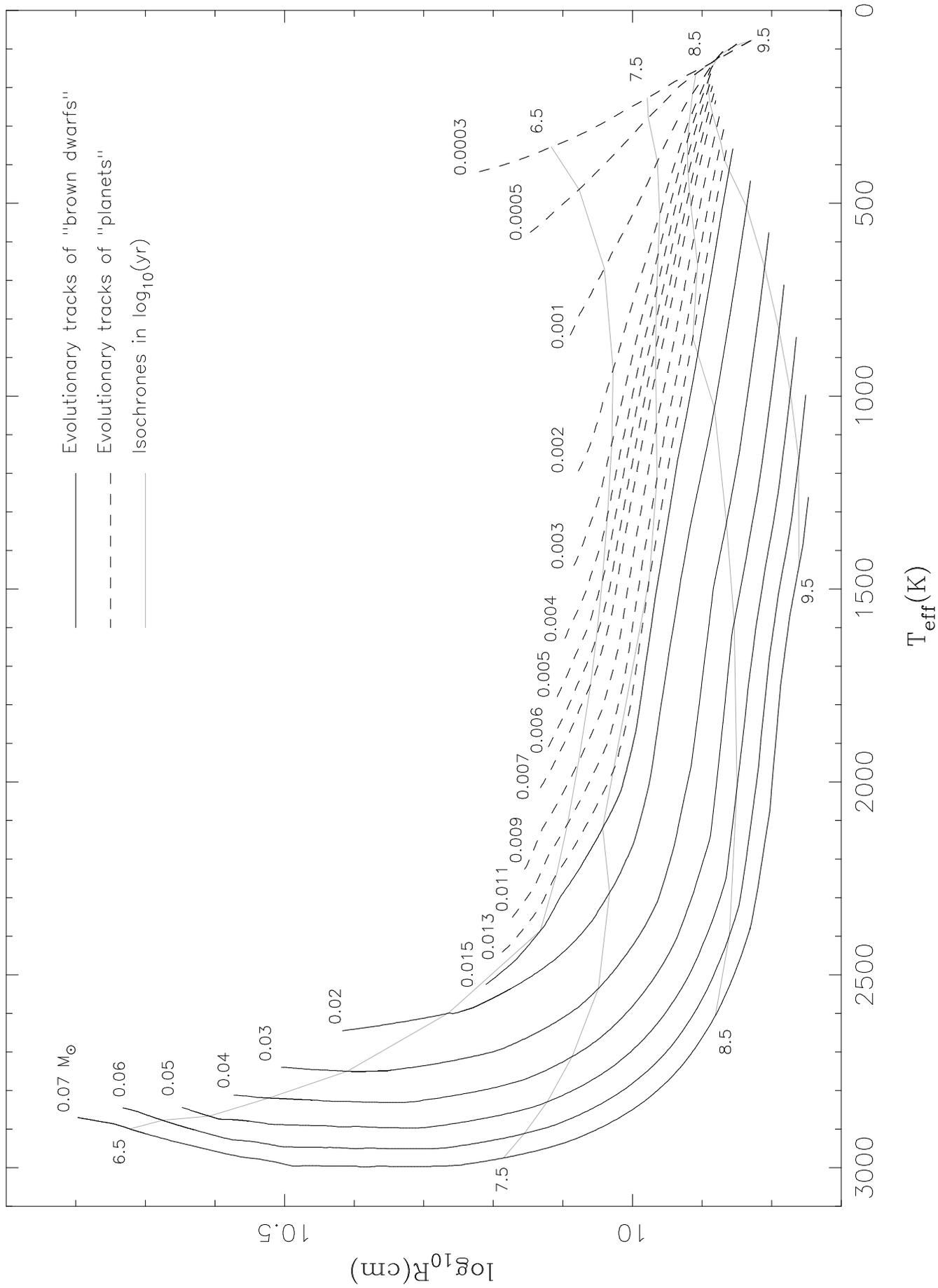
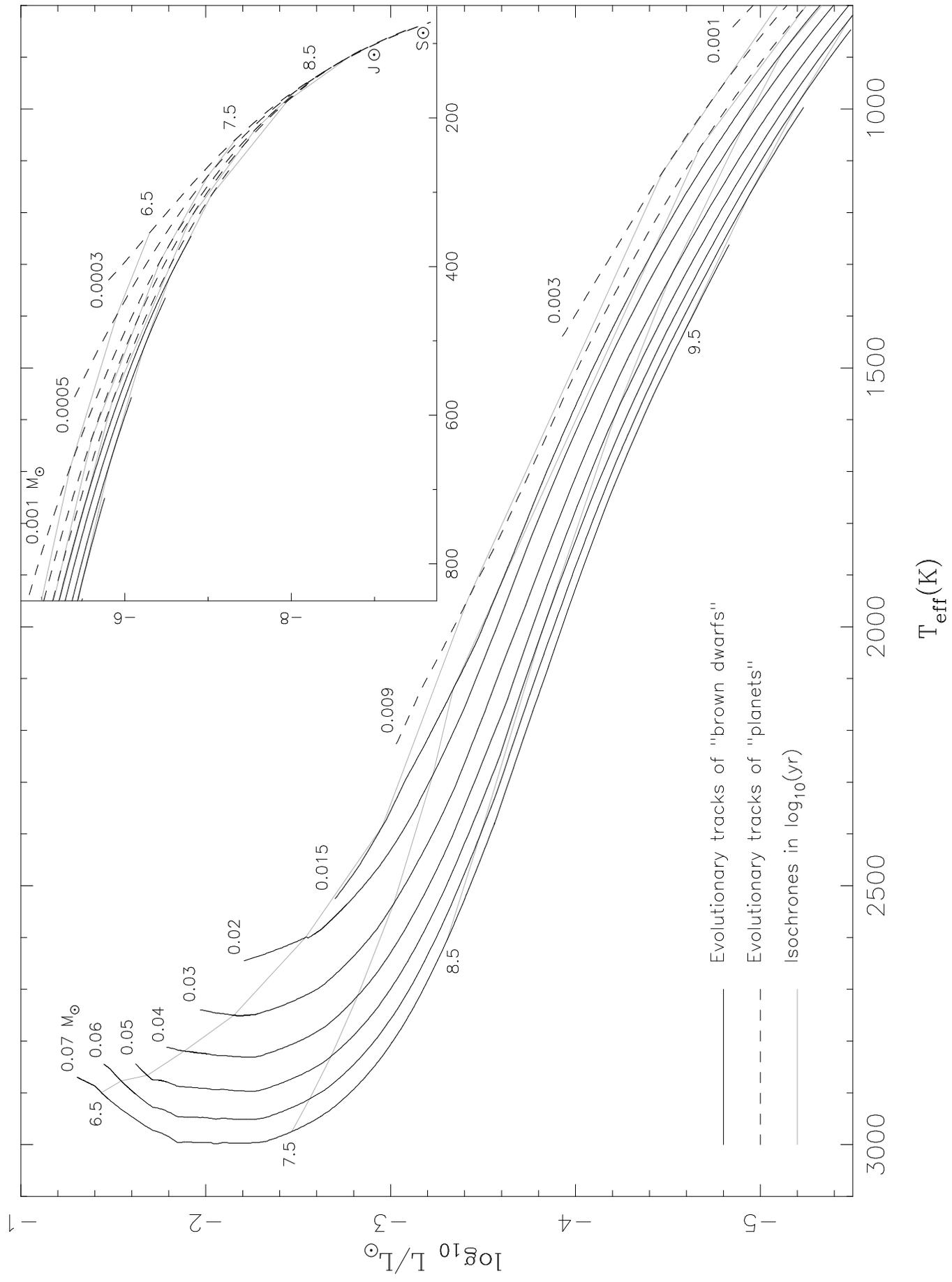
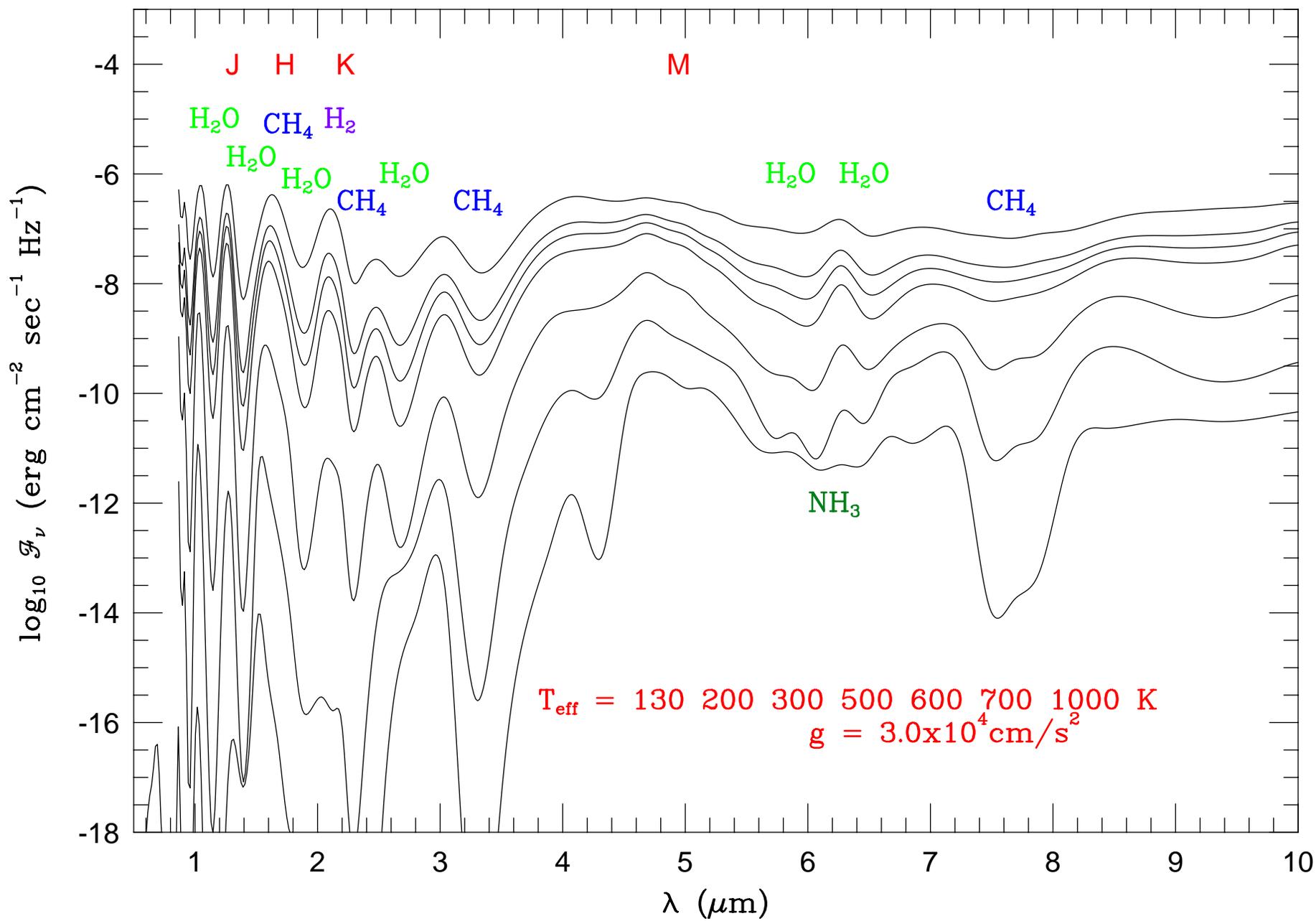
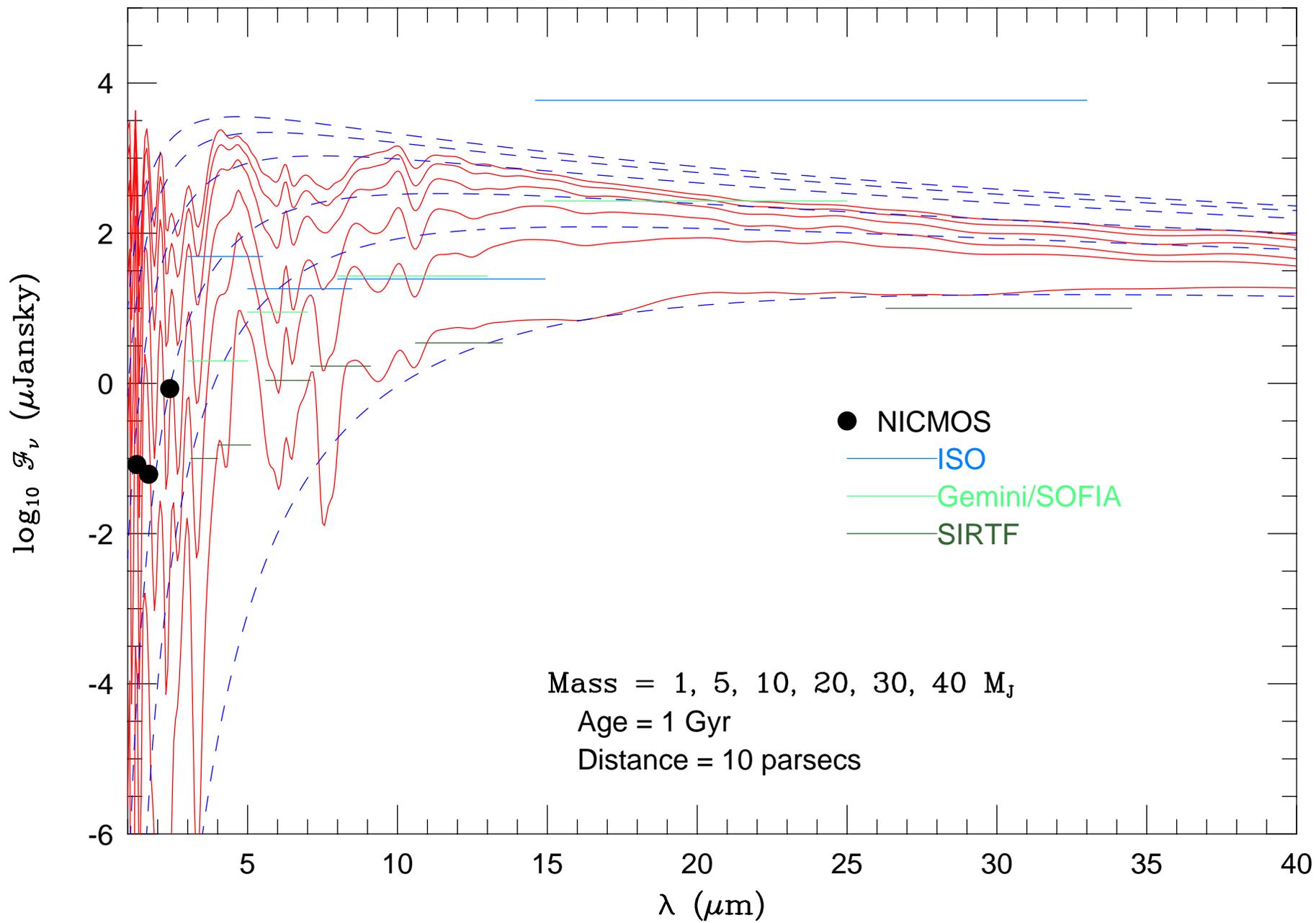
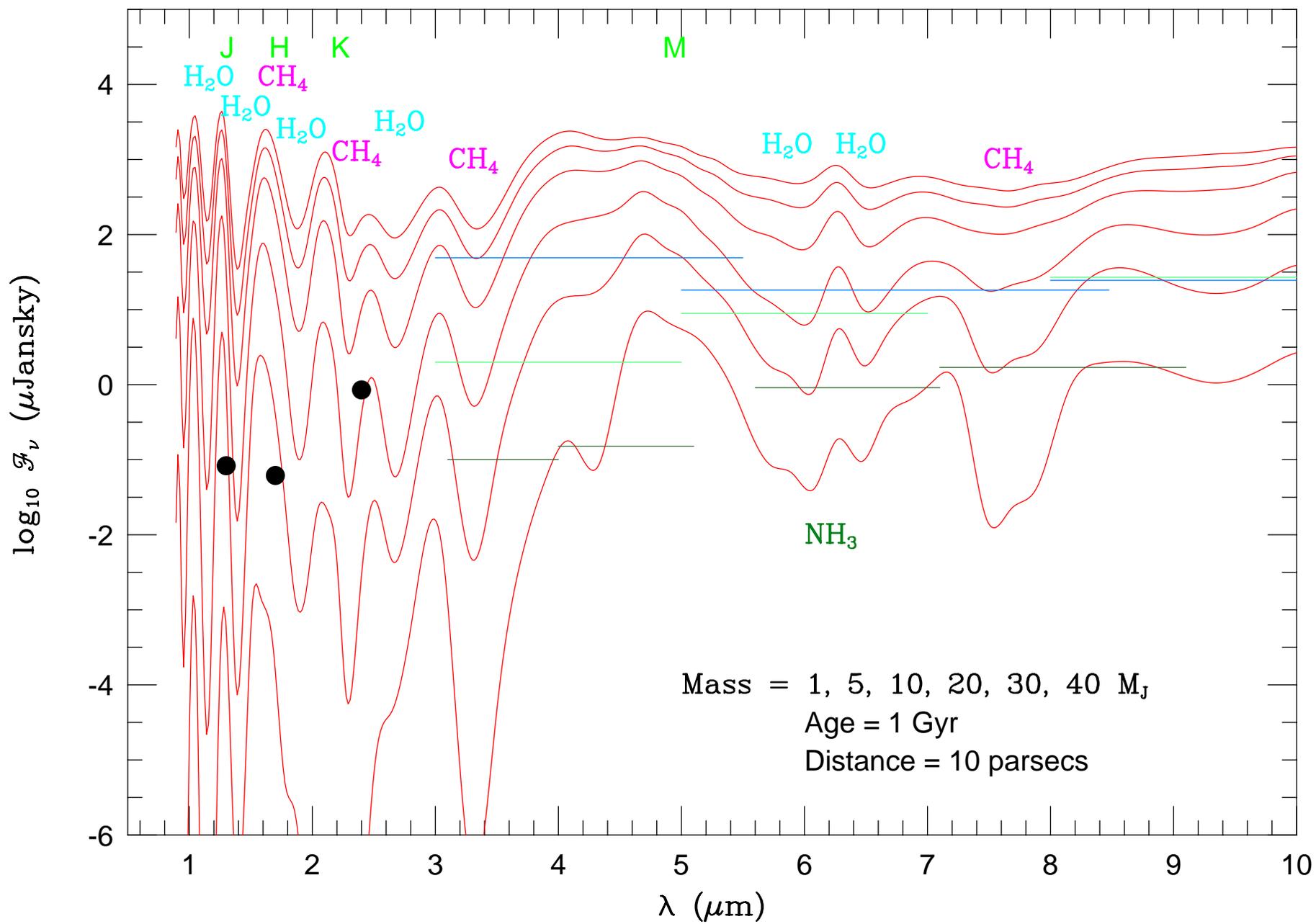
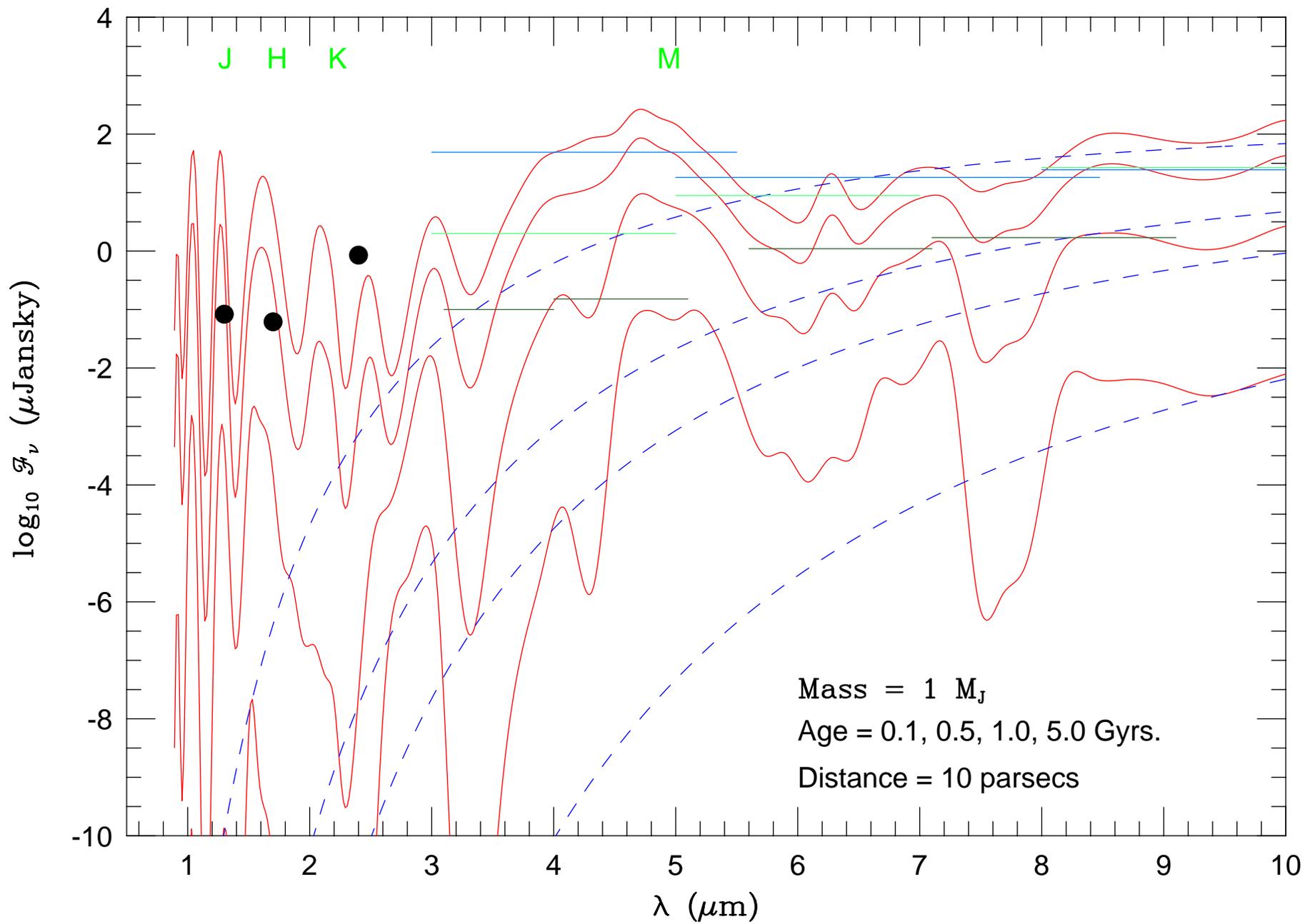
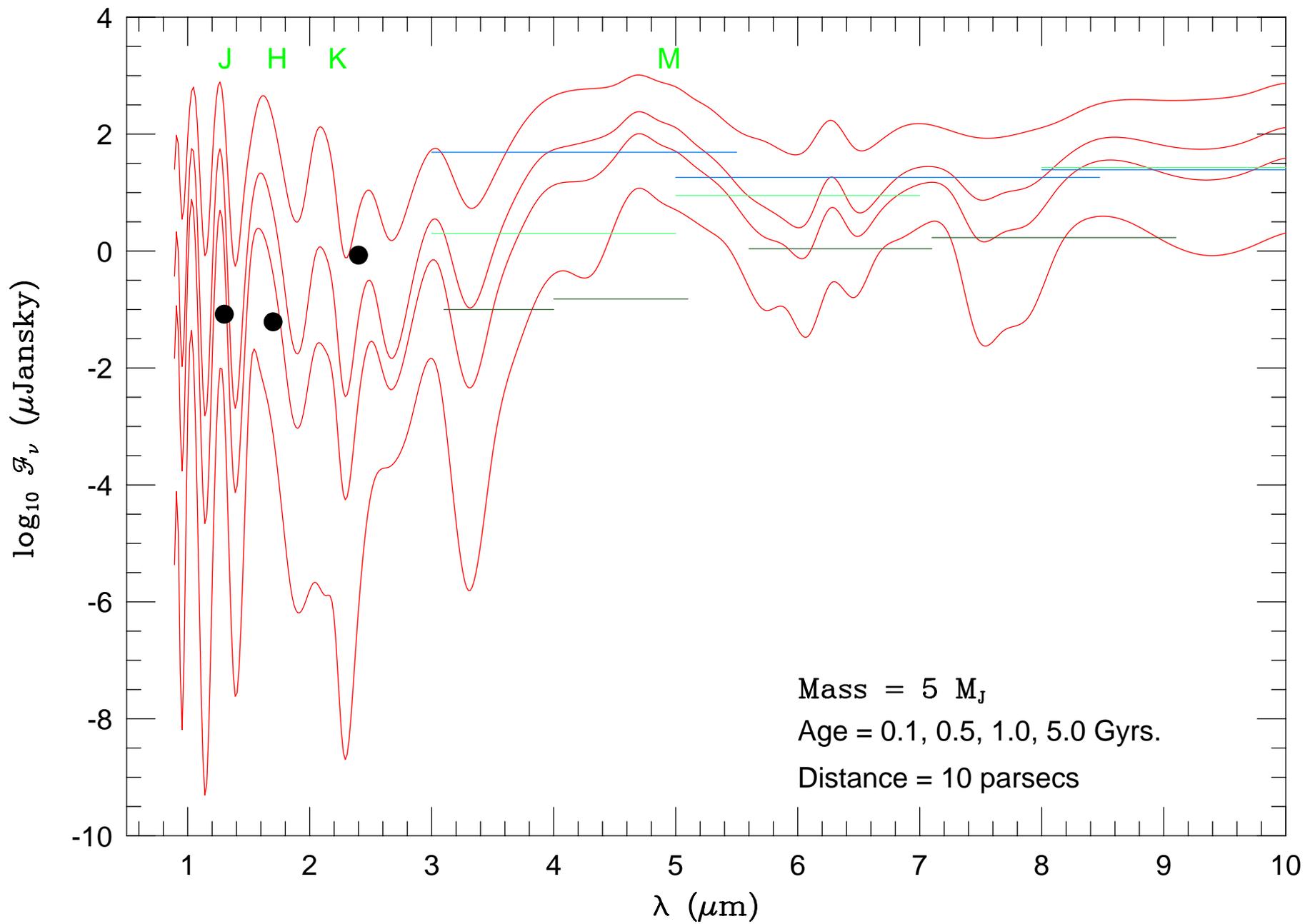
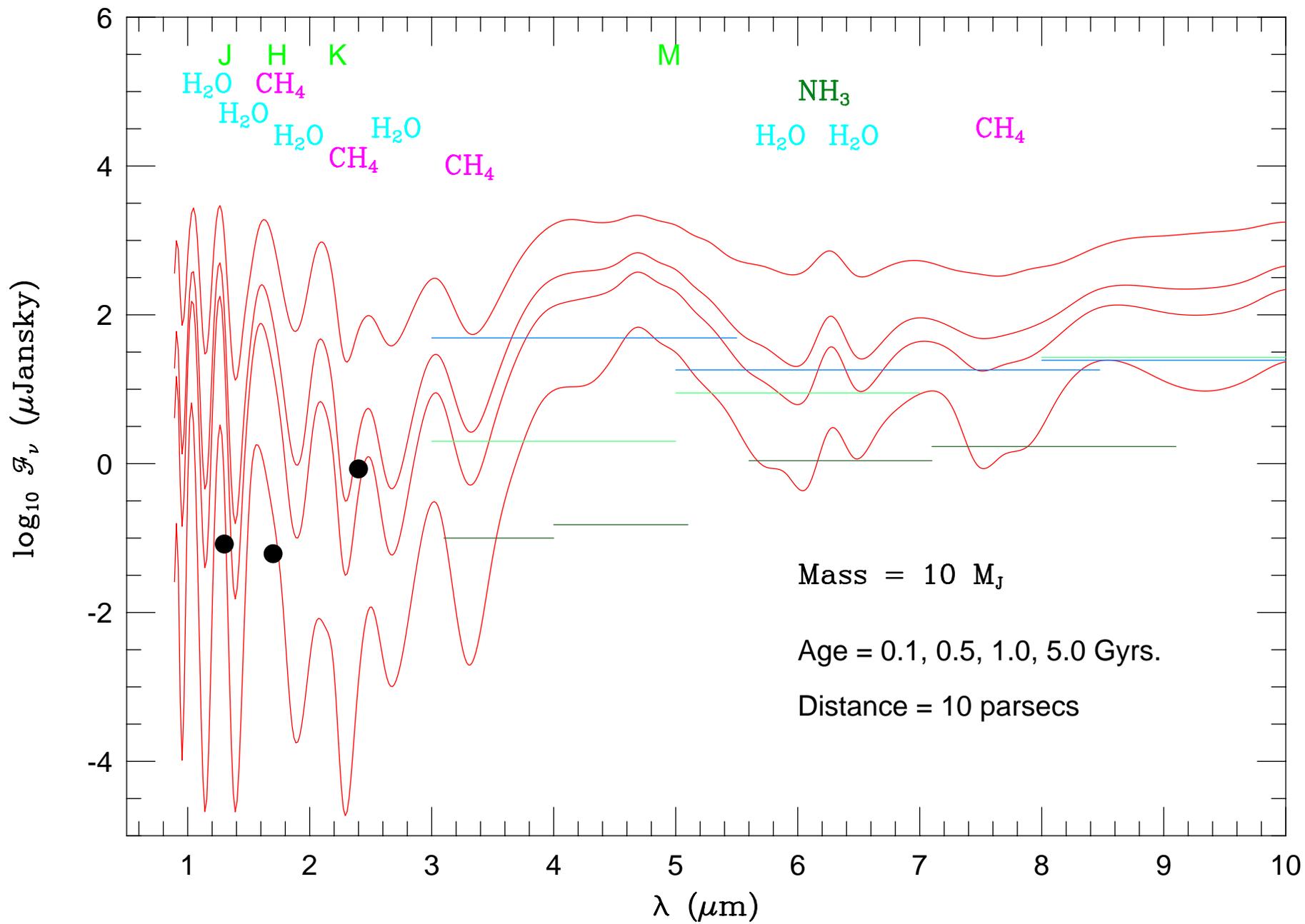
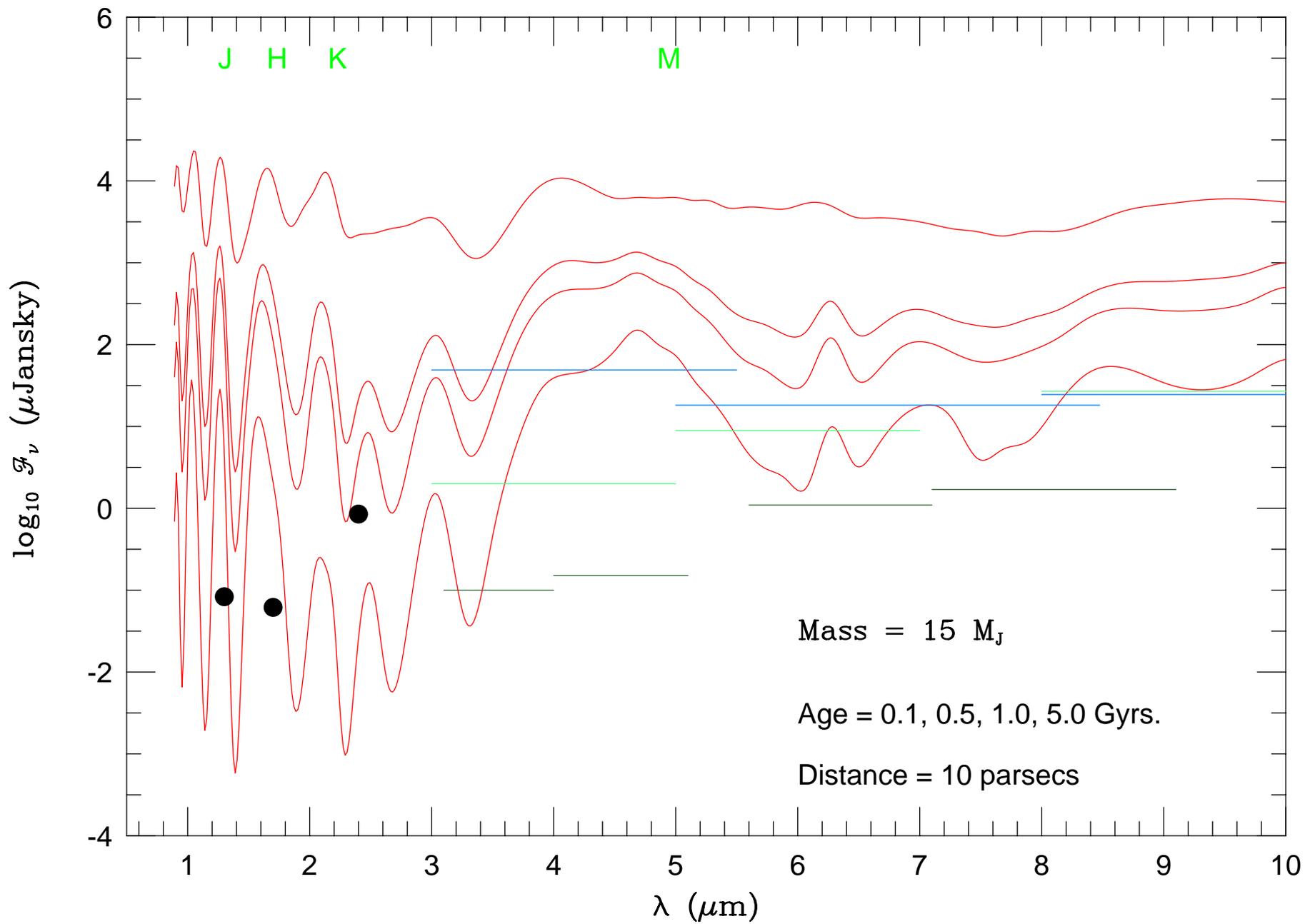
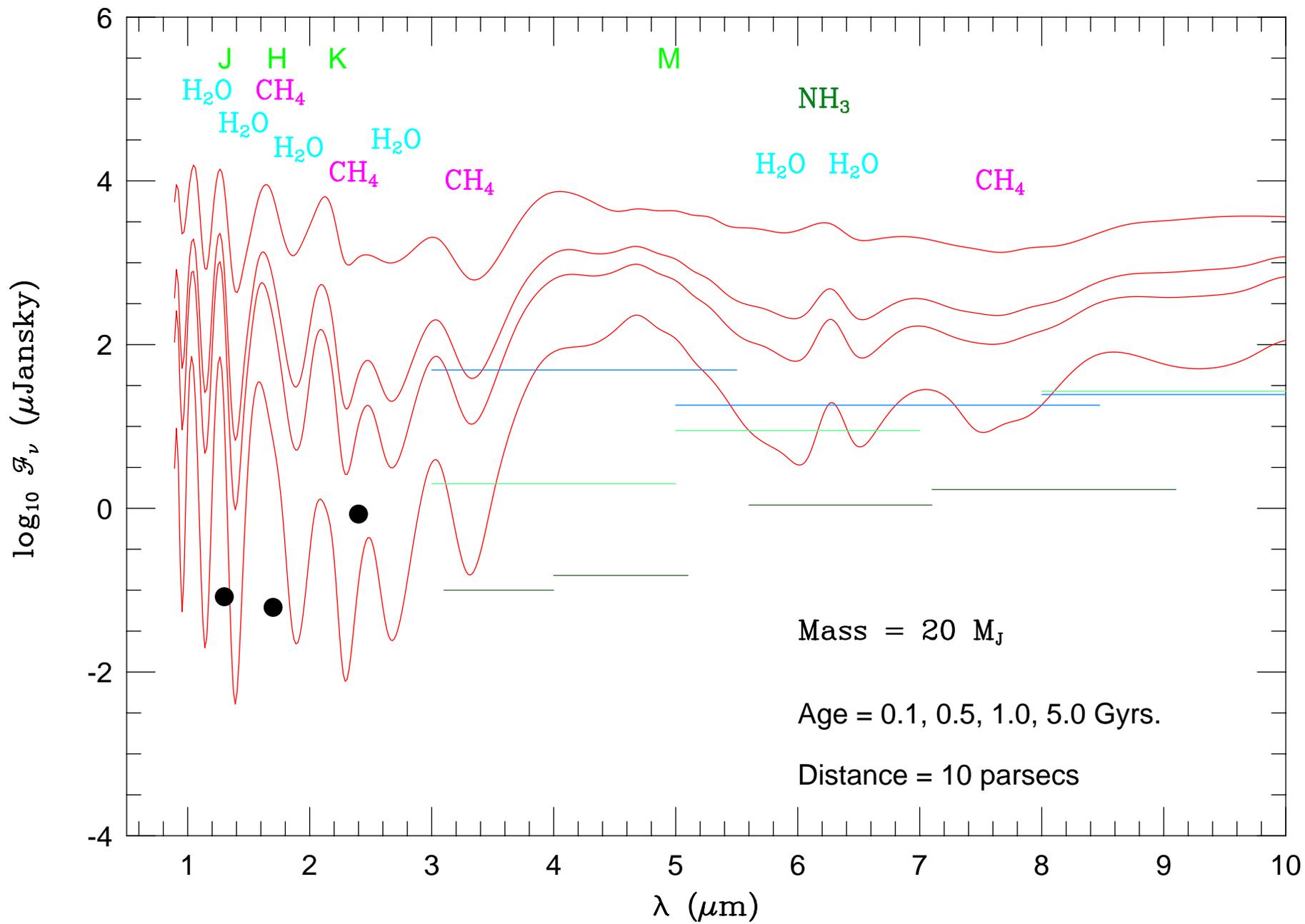
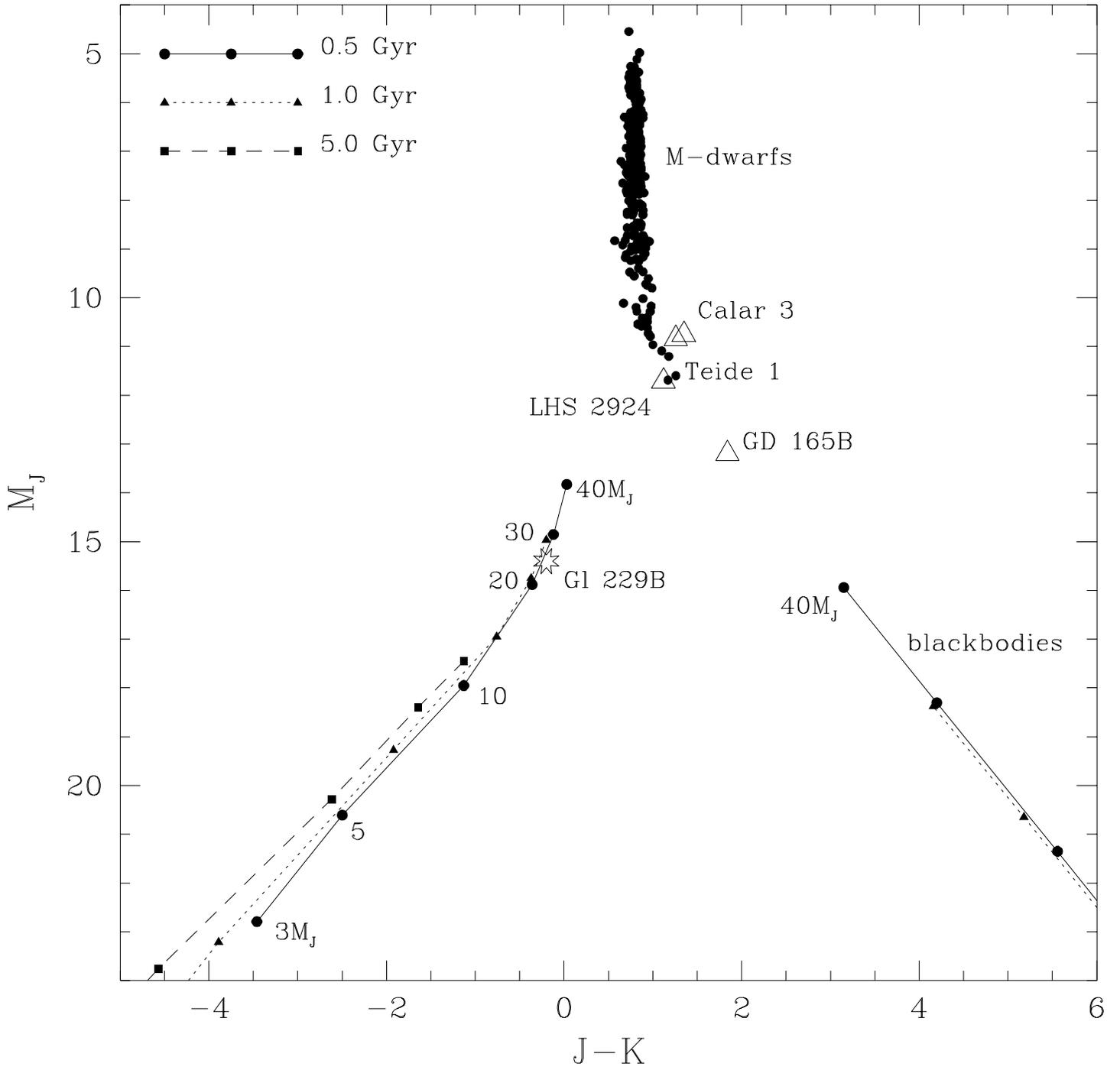
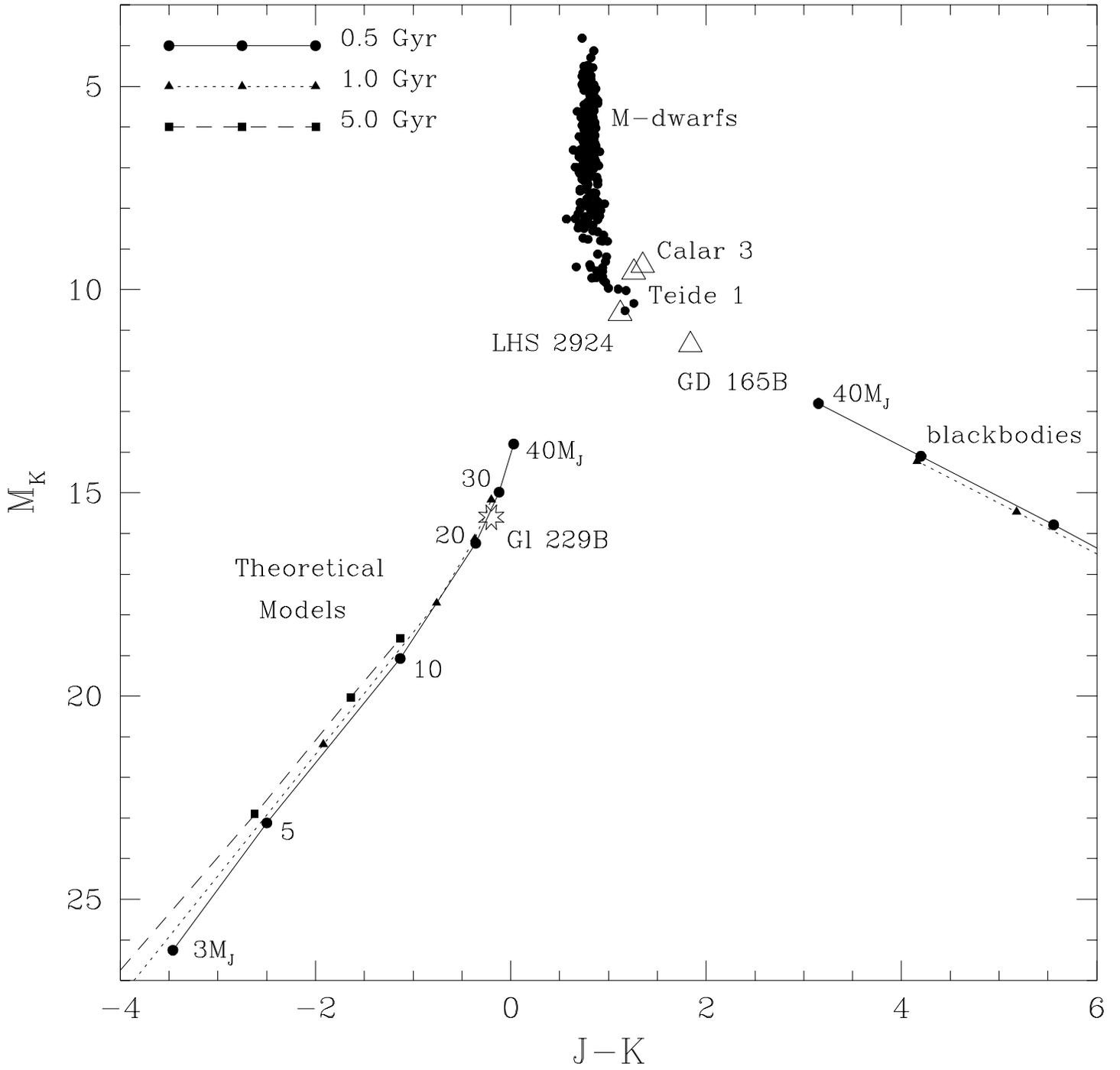
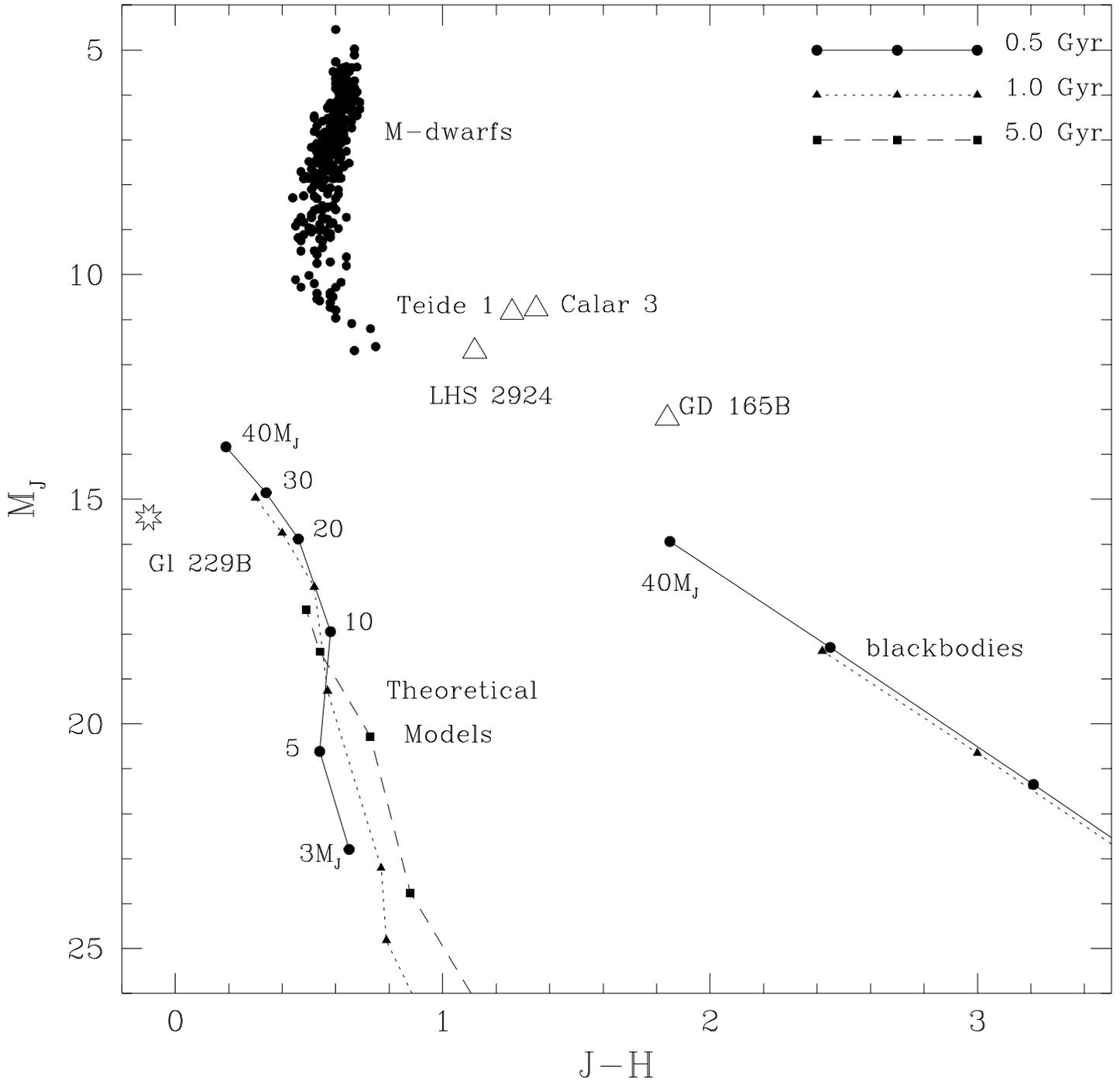
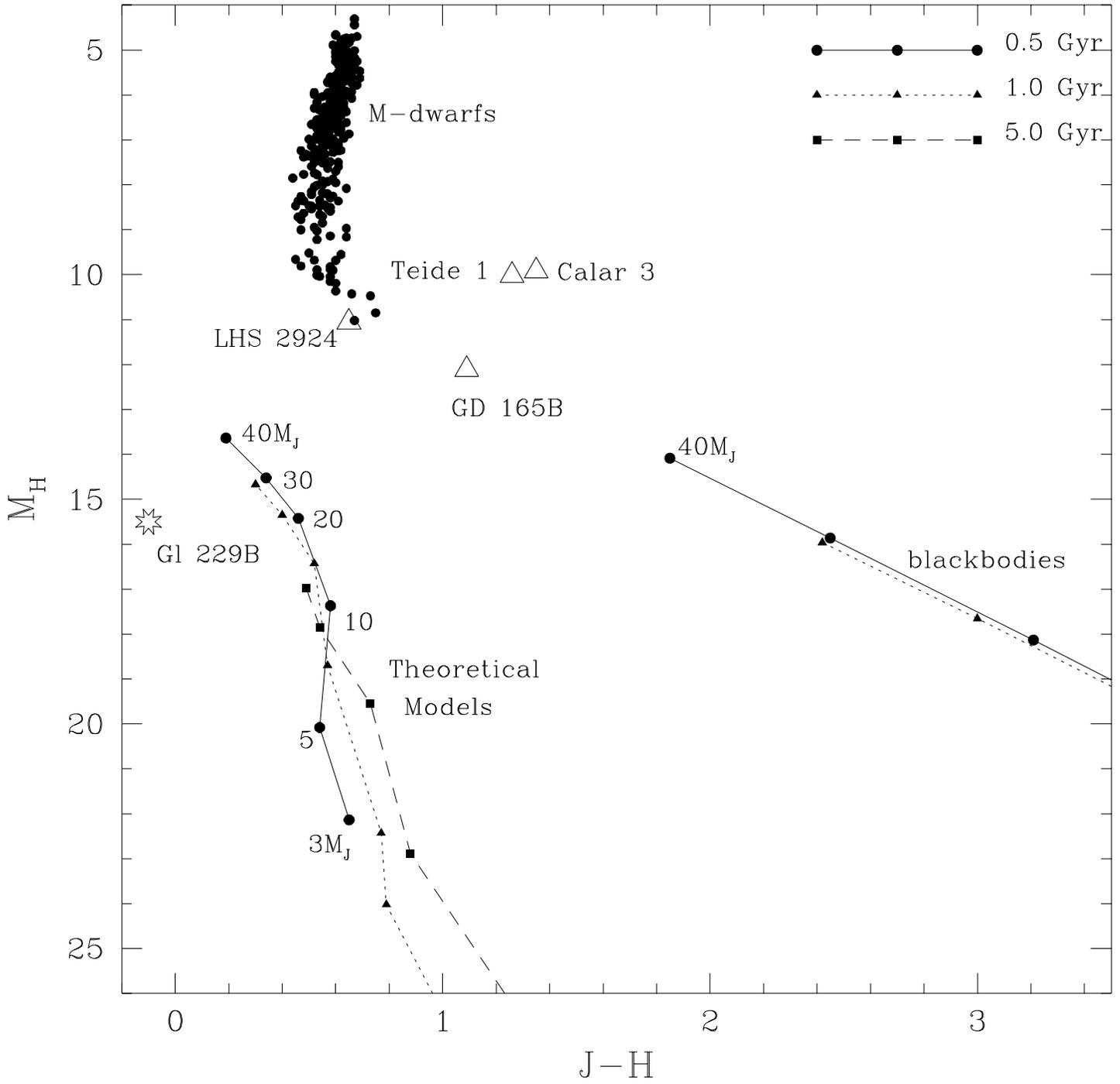
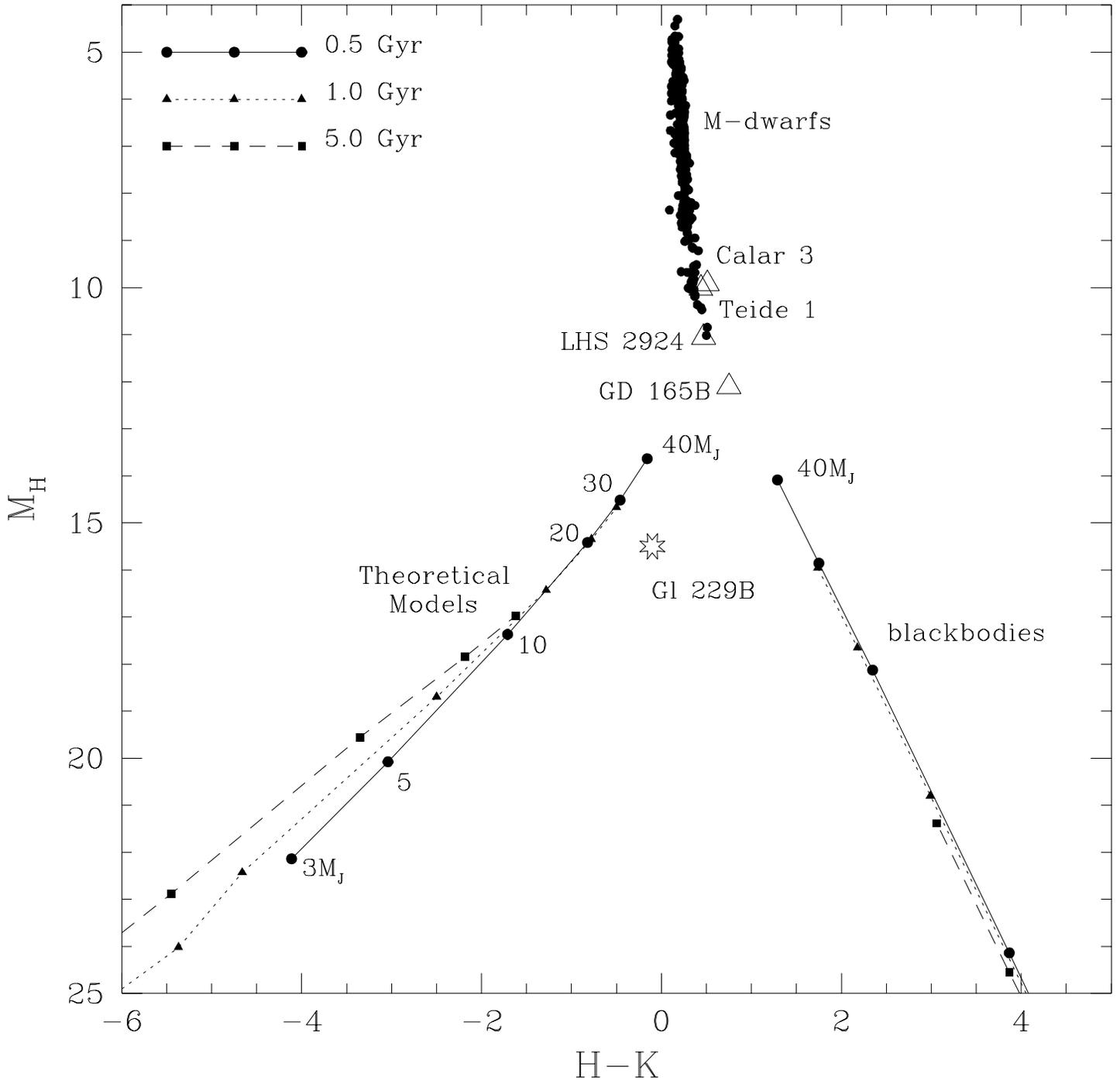
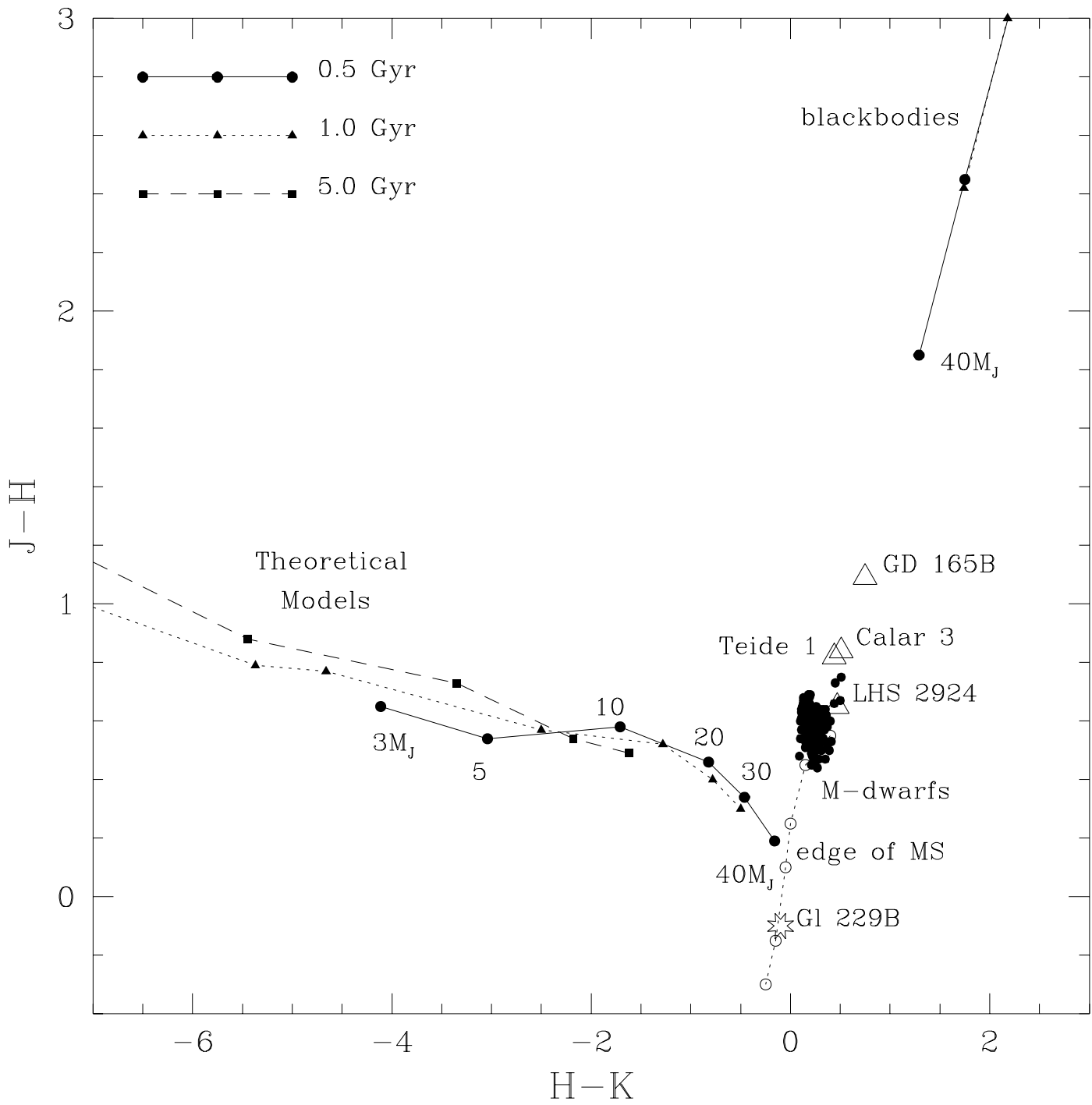
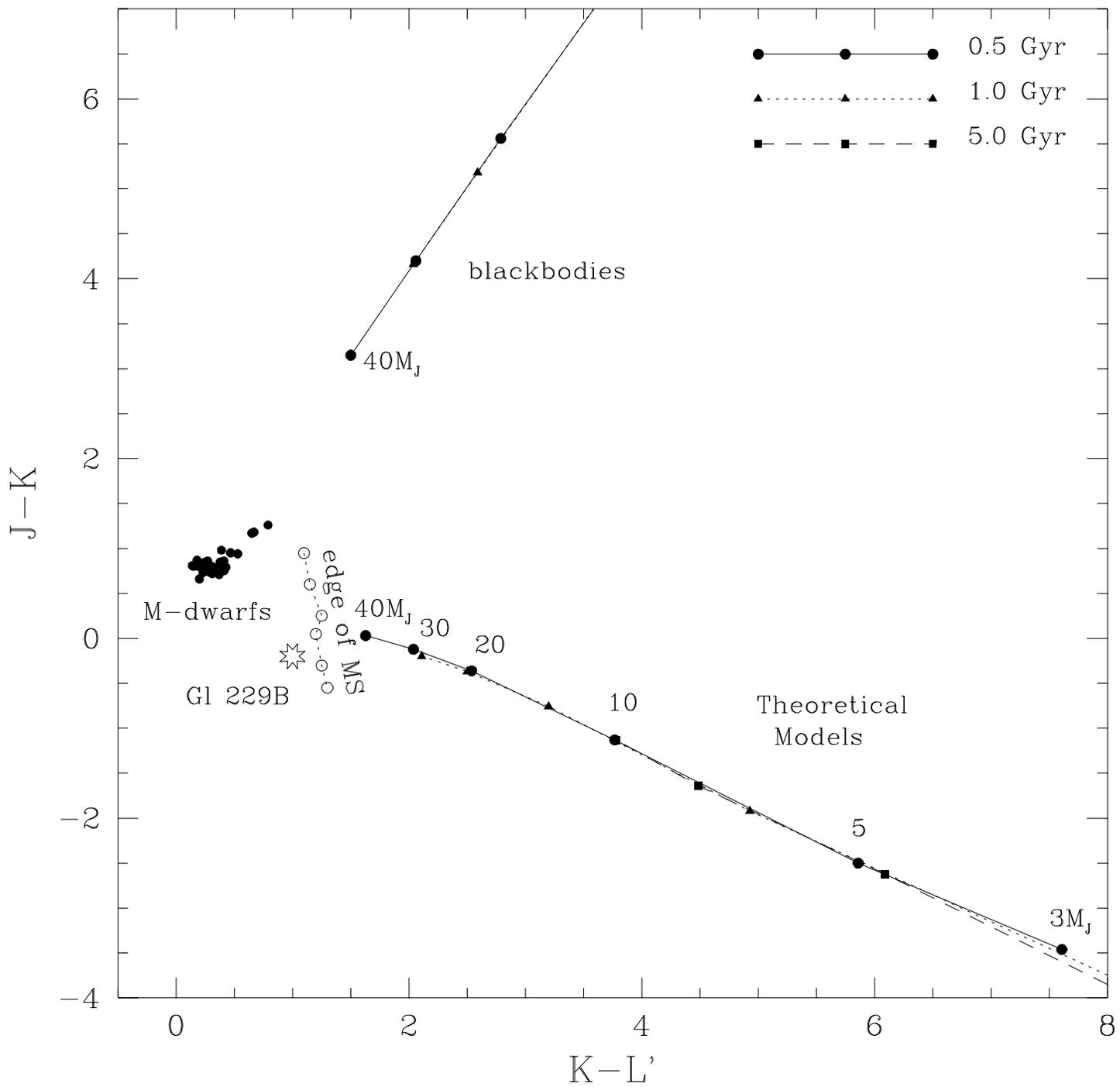
\end{document}